\documentclass{article}

\usepackage{arxiv}

\usepackage[utf8]{inputenc} 
\usepackage[T1]{fontenc}    
\usepackage{url}            
\usepackage{booktabs}       
\usepackage{amsfonts}       
\usepackage{nicefrac}       
\usepackage{microtype}      
\usepackage{lipsum}
\usepackage{verbatim}
\usepackage{graphicx}
\usepackage{footnote}
\usepackage{pifont}
\usepackage{epsfig}
\usepackage{bigints}
\usepackage{pbox}
\usepackage{algorithm}
\usepackage[noend]{algpseudocode}
\usepackage{amsmath}
\usepackage{appendix}
\usepackage{booktabs}
\usepackage{stackengine}
\newcommand\xrowht[2][0]{\addstackgap[.5\dimexpr#2\relax]{\vphantom{#1}}}
\usepackage{dcolumn}
\usepackage[export]{adjustbox}
\usepackage{subcaption}
\usepackage{tikz}
\usepackage{xcolor}
\usepackage{lipsum}
\usepackage{setspace}
\usepackage{mathrsfs}
\usepackage{blindtext}
\usepackage{ragged2e}
\usepackage[makeroom]{cancel}
\usepackage{soul}
\usepackage{float}
\usepackage{multirow}
\usepackage{tabu}
\usepackage{mathtools}
\usepackage{amsthm}
\usepackage{setspace}
\usepackage{hyperref} 
            \hypersetup{backref=true,       
                    pagebackref=true,               
                    hyperindex=true,  
			    hyperfootnotes=false,              
                    colorlinks=true,                
                    breaklinks=true,                
                    urlcolor= black,                
                    linkcolor= blue,                
                    bookmarks=true,                 
                    bookmarksopen=false,
                    citecolor=blue,
                    filecolor=blue,
                    citecolor=blue,
                    linkbordercolor=blue
}
\usepackage{enumitem}
\usepackage{caption}
\makeatletter
\def\@footnotecolor{red}
\define@key{Hyp}{footnotecolor}{%
\HyColor@HyperrefColor{#1}\@footnotecolor%
}
\def\@footnotemark{%
    \leavevmode
    \ifhmode\edef\@x@sf{\the\spacefactor}\nobreak\fi
    \stepcounter{Hfootnote}%
    \global\let\Hy@saved@currentHref\@currentHref
    \hyper@makecurrent{Hfootnote}%
    \global\let\Hy@footnote@currentHref\@currentHref
    \global\let\@currentHref\Hy@saved@currentHref
    \hyper@linkstart{footnote}{\Hy@footnote@currentHref}%
    \@makefnmark
    \hyper@linkend
    \ifhmode\spacefactor\@x@sf\fi
    \relax
 }%
\makeatother
\hypersetup{footnotecolor=black}

\title{A Generalized Bayesian Approach to Model Calibration\thanks{This work was supported by the Center for Complex Systems (CCS) at King Abdulaziz City for Science and Technology (KACST) and the Massachusetts Institute of Technology (MIT).}}

\author{
  Tony Tohme\thanks{Corresponding Author.}, \,Kevin Vanslette, \,Kamal Youcef-Toumi\\
  Department of Mechanical Engineering\\
  Massachusetts Institute of Technology\\
  Cambridge, MA 02139 \\
  \texttt{\{tohme, kvanslet, youcef\}@mit.edu} \\
}
\begin{document}
\maketitle

\begin{abstract}
In model development, model calibration and validation play complementary roles toward learning reliable models. 
In this article, we expand the Bayesian Validation Metric framework to a general calibration and validation framework by inverting the validation mathematics into a generalized Bayesian method for model calibration and regression. We perform Bayesian regression based on a user's definition of model-data agreement. This allows for model selection on any type of data distribution, unlike Bayesian and standard regression techniques, that ``fail" in some cases. We show that our tool is capable of representing and combining least squares, likelihood-based, and Bayesian calibration techniques in a single framework while being able to generalize aspects of these methods. This tool also offers new insights into the interpretation of the predictive envelopes (also known as confidence bands) while giving the analyst more control over these envelopes. We demonstrate the validity of our method by providing three numerical examples to calibrate different models, including a model for energy dissipation in lap joints under impact loading. By calibrating models with respect to the validation metrics one desires a model to ultimately pass, reliability and safety metrics may be integrated into and automatically adopted by the model in the calibration phase.
\end{abstract}


\section{Introduction}

The increasing complexity of engineering systems over the years has required a thorough understanding and assessment of their behaviour for their development, and the need for a framework that is capable of quantifying their reliability has become essential. One of the main objectives of reliability engineering is to identify and reduce the likelihood (or frequency) of failures. Given that most of the systems that constitute an essential part of our life (e.g. automotive, medical, energy, nuclear, etc.) are safety-critical systems, any failure in those systems can be harmful to people and the environment \cite{kabir2019}. Hence, reliability, safety, and risk analysis \cite{zio2016, zio2018} are fundamental for designing, modeling, and testing any modern engineering system \cite{zio2007}, and essential for risk-informed decision-making \cite{penttinen2019}.\\

Parametric modeling is an integral part of reliability engineering and plays a critical role in designing, analysing and predicting the behavior of complex engineering systems \cite{vanderhorn2018}. When performing model-based reliability analysis, the uncertainty in the model parameters greatly affects the accuracy of the analysis. This uncertainty can be reduced by merging prior knowledge about the model parameters with available data describing the relation between the system inputs and outputs. Uncertainty can be aleatoric (natural variability) and/or epistemic (lack of knowledge) \cite{vanderhorn2018, sankararaman2011, nannapaneni2016}. In this paper, we are interested in formulating a probabilistic framework -- by expanding the Bayesian Validation Metric (BVM) framework \cite{Vanslette2019} -- for updating the model parameters as well as estimating and quantifying their uncertainty and the propagation of this uncertainty to the model response \cite{haverkort1995}. Our framework and methodology can be applied to a broad range of reliability engineering problems (e.g. the energy dissipation model \cite{Rebba, Sankararaman2}, as illustrated in Section \ref{example3}, heat transfer models for fire insulation panels \cite{wagner2020}, energy models \cite{chong2019}, dynamic thermal models \cite{raillon2018}, etc.) and deals with uncertainty through the use of confidence intervals and probability distributions \cite{yin2001}.\\

Engineering systems are often represented and described by computational models in order to make predictions about the behavior of the system. Most of the time, models possess parameters that cannot be directly measured, and instead they are inferred based on experimental data of relevant inputs and outputs, in a process known as model calibration \cite{Mullins, Trucano, Lee, Wu}. Model calibration is the process of estimating and adjusting model parameters to obtain a model representation of the system (or data) of interest while satisfying a specific criterion (objective function). In model development, model calibration comes in the stage prior to the validation stage \cite{SANKARARAMAN2015194}, and it usually consists of estimating model parameters given a set of observed input-output data. Then, validation is performed on a different independent data set called the validation data set. In Bayesian calibration, one generates a posterior distribution for the model parameters given some prior distribution of those parameters and the data available for calibration. In what follows, we refer to Bayesian calibration as Bayesian regression.\\

Representing and understanding data through learning models (by estimating their parameters) has always been a central problem in engineering and science.
Least squares (or standard regression) \cite{wild1989nonlinear}, likelihood-based \cite{edwards1984likelihood, pawitan2001all}, and Bayesian regression methods \cite{leonard2001bayesian, lee1997bayesian, malinverno2004expanded, Park, epsilonmetric} are often used for model parameter estimation. Nonprobabilistic methods, such as parametric model regression, nonparametric neural networks, and support vector machines (SVM) \cite{Bishop2006} are able to tackle these types of problems efficiently.  In Bayesian probability theory \cite{Mackaybook,Sivia,mhalgo}, Bayesian model testing and maximum likelihood methods provide probabilistic features (i.e. mean, covariance, distribution) for the parameters we aim to estimate, based on prior knowledge (i.e. prior distribution) and the uncertainty of the data. Bayesian model testing, which uses Bayesian parameter regression, was shown to be successful for signal detection, light sensor characterization \cite{knuth}, exoplanet detection \cite{Placek1}, extra-solar planet detection \cite{Placek}, laser peening process \cite{Park}, time series \cite{timeseries}, astronomical data analyses \cite{FH08}, and cosmology and particle physics \cite{FH09}. \\
    
\begin{figure} [H]
\centering
\begin{subfigure}{8cm}
\includegraphics[width=8cm]{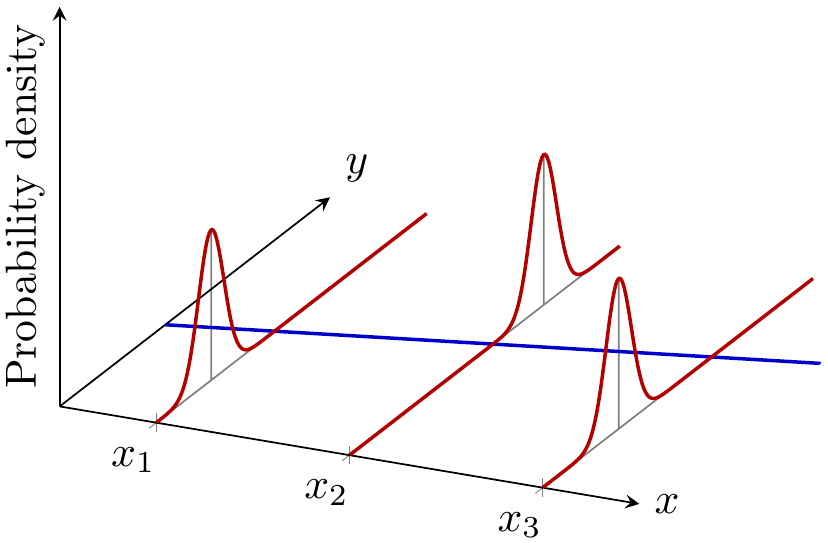}
\caption{Bayesian regression works.\label{fig1a}}
\end{subfigure}
\:\:
\begin{subfigure}{8cm}
\includegraphics[width=8cm]{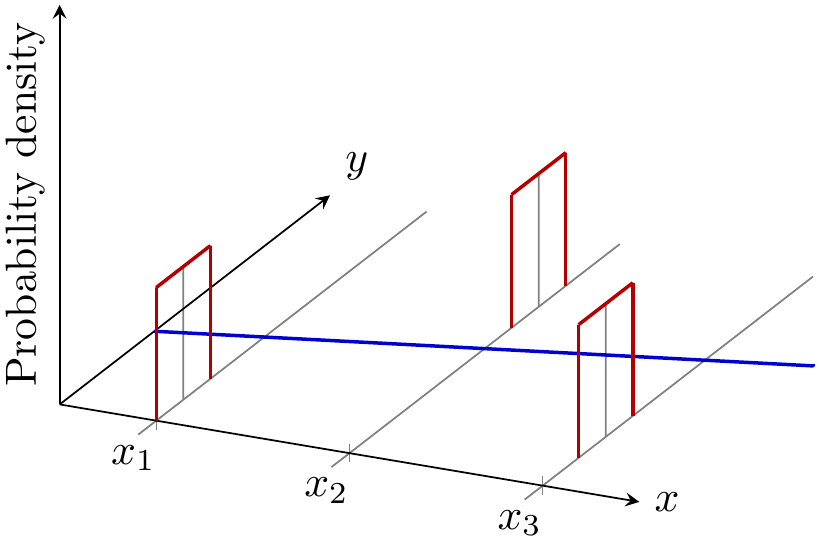}
\caption{Bayesian regression fails.\label{fig1b}}
\end{subfigure}
\caption{Illustrative example of theoretical success and failure cases of Bayesian regression. In blue is a deterministic linear model ($y=ax+b$) and in red are the data probability distributions that may come from epistemic uncertainty, (constrained) aleatoric uncertainty, or from their combination.
\label{fig1}}
\end{figure}

We believe that the efficacy of parametric Bayesian regression, likelihood-based, and standard regression methods can be improved. 
Bayesian regression methods calculate the Bayesian evidence, which is the probability the model could have produced the observed, usually over noisy or uncertain, data. If this probability is nonzero, one can proceed to calculate posterior model parameter probabilities using Bayes' Theorem. In practice, there are models and parameters that may be of interest to the user that Bayesian regression fails to regress and produce posterior parameter distributions -- Figure \ref{fig1}. For some of the instances that Bayesian regression fails to provide a solution, standard regression methods may actually succeed, but usually with some measure of expected error. How this error can translate into parameter and model uncertainty in the presence of certain or uncertain data is a problem that is largely omitted in the literature except for a few analytic cases. In addition, standard regression methods cannot be used for model selection in which one could easily include their prior knowledge in a principled way.

\indent Figure \ref{fig1a} shows normally distributed data (infinite tails data distribution). In this case, parametric Bayesian regression finds a linear model that sits in low probability regions of the data. Figure \ref{fig1b} shows uniformly distributed data (truncated data distribution). In this case, Bayesian regression cannot find a linear model solution because no linear model can pass through each data distribution simultaneously -- the model given the data is regarded as impossible. Standard regression methods can provide linear model solutions here despite the model lying in a zero probability region of the data. Although this solution may be considered ``wrong" because it is not supported by the data, it successfully provides useful information to the modeler (an increasing trend). The fact that, for the same model, the solution given by the calibration method can differ from method to method supports the search for a framework for their joint representation so they may be compared more concretely.\\

The Bayesian Validation Metric (BVM) was shown to be a general model validation and testing tool in \cite{Vanslette2019}. The BVM is capable of representing all of the standard validation metrics (square error \cite{epsilonmetric}, reliability \cite{Rebba}, improved reliability \cite{Sankararaman2}, probability of agreement \cite{Stevens}, frequentist \cite{Oberkampf}, area \cite{Wu}, statistical hypothesis testing \cite{Zhang}, Bayesian model testing \cite{Sivia,Placek,Zhang,Sankararaman}, etc.) as special cases. Using BVM model testing, the BVM selects models according to whatever definition of model-data agreement the modeler might find useful. It was found that the BVM is able to generalize the Bayesian model testing framework, which allowed the problem of model validation to be expressed in a single framework.\\ 
    
In this article we represent least squares, likelihood-based, and Bayesian regression (or calibration) methods by expanding the general validation framework (BVM) into a general calibration and validation framework. Our method uses the BVM to guide the regression of a model in a flexible way. Several of our examples use generalizations of the improved reliability metric and thus reliability is automatically taken into account in our model solutions. Our method gives us better control over the predictive envelopes of the model under question, which can be used to improve model reliability and safety. By learning model parameters with the BVM, we are able to estimate and construct model parameters distributions for any type of data distribution (Gaussian, Uniform, Completely Certain, etc.), which addresses the concerns raised in Figure \ref{fig1}. 
This construction gives us additional insight into the meaning of the predictive envelopes of Bayesian regression methods.\\
    
We have found that a subset of our method shares mathematical features with Approximate Bayesian Computation (ABC) methods, which are also known as likelihood-free techniques \cite{Beaumont2025, ABCMethods}. ABC methods are used strictly as an approximation method for nearly computationally intractable likelihoods in Bayesian regression. While our method gains this feature in some cases, our method's intention is not to approximate Bayesian regression, but instead to generalize it for the purpose of robust and flexible model calibration. \\

Our method is able to regress models over a multitude of different data distributions by using likelihoods that are modified by user's choice of a useful definition of agreement between the data and the model -- leaning on the BVM formalism. This ``choice" allows the user to program safety requirements into the model learning process if they desire. The nature of the BVM formalism forces one to express the model and data assumptions explicitly and thus, our method leads to improved model transparency and safety. In our examples we show how such a procedure leads to a model that better represents the uncertain data at hand than Bayesian and standard regression techniques. This naturally improves the model's reliability and safety in the presence of uncertain data. \\ 
    
The remainder of the article is organized as follows. In Section \ref{background}, we discuss the research contribution and motivation, review Bayesian regression and model testing as well as the Bayesian Validation Metric (BVM). We then move to Section \ref{bvmregression} where we derive our theoretical solutions for BVM regression and model selection (or learning), under different types of data distributions (or uncertainty) and according to user-specified definitions of model-data agreement. In Section \ref{simulation}, we illustrate the proposed method and provide three numerical examples for model calibration including a model used to predict the energy dissipation in lap joints under dynamic loading. 
    
\section{Contribution and Motivation\label{background}}

In this section, we will outline the research contribution and motivation. We will also introduce the notation adopted in this paper, review Bayesian regression methods, discuss their advantages and disadvantages, and review the Bayesian Validation Metric (BVM).
\subsection{Contribution}
In our earlier work \cite{Vanslette2019}, we introduced and constructed the Bayesian Validation Metric (BVM) as a general model validation and testing tool. The BVM was shown to represent and generalize all of the well-known validation metrics (e.g. reliability metric, etc.). The BVM was also shown to generalize Bayesian Model Testing and allow for model selection based on arbitrary definitions of model-data agreement. In this paper, we extend this work and present further development of the BVM framework that results in generalizing Bayesian methods for regression and model learning. The resulting framework can then be used in both the calibration and validation stages of model development. Estimating and quantifying predictive uncertainty in least squares (or standard regression) \cite{wild1989nonlinear}, likelihood-based \cite{edwards1984likelihood, pawitan2001all}, and Bayesian regression \cite{leonard2001bayesian, lee1997bayesian, malinverno2004expanded, Park, epsilonmetric} methods is a very challenging task. In addition, these methods face several problems (and fail) when the data they are dealing with have bounded probability distributions or are deterministic (completely certain). \\

In this paper, we show how our extended BVM framework succeeds and outperforms those methods for any type of data uncertainty (or distribution). In addition, our method is successful when it comes to predictive uncertainty estimation, as we will see in our examples. Given the flexibility that the BVM framework provides when defining the agreement between the model outputs and the observed data, we show how this flexibility translates into more freedom and control over the predictive envelopes resulting from the posterior distributions of the model parameters. Finally, our method is very simple to implement, requiring few modifications to standard Markov Chain Monte Carlo (MCMC) methods or any Bayesian inference tool that calculates the evidence and produces posterior samples (as we will see, the only task required is to come up with or derive the appropriate likelihood function to be fed into the MCMC algorithm). These contributions are formulated in Sections \ref{bvmregression} and \ref{simulation}.

\subsection{Notation}\label{section2.0}

We start by introducing the notation and language used in this paper. For an input $x$, we let $\hat{y}$ denote the output of a model $M$, i.e. $\hat{y} = M(x; \vec{\alpha})$, where $\vec{\alpha} = (\alpha_1, \hdots, \alpha_m)$ represents the vector of model parameters we wish to estimate. We let $X=(x_1,...,x_n)$ denote a set of inputs, $\hat{Y}=(\hat{y}_1,...,\hat{y}_n)$ denote a set of model outputs, i.e. $\hat{Y} = M(X; \vec{\alpha})$, $y$ denote a data point (or observed data), and $Y$ denote a set of data points. We let $\mathcal{L}$ denote the likelihood, $\mathcal{Z}$ denote the marginal likelihood (or model evidence), and $A$ denote model-data \emph{agreement}.

\subsection{Bayesian Regression and Model Testing\label{section2.2}}

In this section, we present Bayesian regression and Bayesian model testing (BMT) while introducing some probability notations to be used throughout this paper. In Bayesian regression, rather than preforming regression to learn the model parameters, one performs regression to learn the posterior probability distribution of the model parameters. That is, one estimates a set of parameters $\vec{\alpha}$ in a model (or hypothesis) $M \equiv M(x;\vec{\alpha})$ for the data $D$ (where $x$ is the model input).
The defining equation of Bayesian regression is the learning of the posterior parameter distribution from the prior via Bayes' Rule, \vspace{-15pt}\\
\begin{align}
\rho(\vec{\alpha}|M)\stackrel{*}{\longrightarrow}\rho(\vec{\alpha}|D,M) = \displaystyle \frac{\rho(D|\vec{\alpha},M)\,\rho(\vec{\alpha}|M)}{\rho(D|M)},\label{BayesianRegression}
\end{align}
where, for reasons that will become obvious, we have borrowed the more explicit notation from \cite{Vanslette2019}. In Bayesian regression and model testing, these probabilities are named as follows:\vspace{5pt}\\
\indent\hspace{2pt}$\rho(\vec{\alpha}|D,M) \hspace{0.7pt}\equiv \mathcal{P}(\vec{\alpha})$ is the posterior probability of the parameter,\vspace{2pt}\\
\indent\hspace{2pt}$\rho(D|\vec{\alpha},M) \hspace{1 pt}\equiv \mathcal{L}(\vec{\alpha})$ is the likelihood function,\vspace{2pt}\\
\indent\hspace{0pt}$\rho(\vec{\alpha}|M) \hspace{16.5pt}\equiv \pi(\vec{\alpha})$ is the prior probability,\vspace{2pt}\\
\indent\hspace{0pt}$\rho(D|M) \hspace{14.5pt}\equiv \mathcal{Z}$ is the marginal likelihood or Bayesian evidence.\vspace{10pt}\\
\noindent After learning the posterior distribution of the model parameters (given $\vec{\alpha}$ is the vector of parameters), we can evaluate the predictive distribution defined by:\vspace{-15pt}\\ 
\begin{align}
\rho(\hat{y}|D,M) = \displaystyle\int_{\vec{\alpha}} \rho(\hat{y}|\vec{\alpha},D,M)\cdot\rho(\vec{\alpha}|D,M) \:d\vec{\alpha}.\label{BMTPredict}
\end{align}
To perform Bayesian regression, one must calculate the Bayesian evidence, which is the marginal likelihood over $\vec{\alpha}$,\vspace{-15pt}\\
\begin{align}
\mathcal{Z} = \rho(D|M) = \displaystyle\int_{\vec{\alpha}}\underbrace{\displaystyle  \rho(D|\vec{\alpha}, M)}_{\displaystyle \mathcal{L(\vec{\alpha})}}\cdot\underbrace{\vphantom{\displaystyle \rho(D|\vec{\alpha}, M)}\rho(\vec{\alpha}|M)\,d\vec{\alpha}}_{\displaystyle \mathcal{\pi(\vec{\alpha})}\,d\vec{\alpha}}.\label{BayesianEvidence}
\end{align}
After performing regression and solving for the model parameters' values, rather than selecting the model with the lowest  estimated generalization error as is done in standard regression, one instead uses BMT to select the model with the highest probability given the data. That is, for two Bayesian regressed models $M_1$ and $M_2$, BMT uses the Bayes ratio, $R$, and rank the data-informed posterior model probabilities. It can be expressed in several ways using Bayes Rule,\vspace{12pt}\\
\centerline{$\displaystyle R\equiv \frac{p(M_1|D)}{p(M_2|D)} = \frac{\rho(D|M_1)\,p(M_1)}{\rho(D|M_2)\,p(M_2)} = \frac{\mathcal{Z}_1}{\mathcal{Z}_2}\frac{p(M_1)}{p(M_2)}.$}\vspace{12pt}\\
If there is no reason to suspect that one model is more probable than another prior to observing the data, we may set the ratio of the prior probabilities of the model $p(M_1)/p(M_2) = 1$, \emph{a priori}. In this case one gets,\vspace{10pt}\\
\centerline{$\displaystyle R\rightarrow \displaystyle \frac{\mathcal{Z}_1}{\mathcal{Z}_2} \equiv K, $}\vspace{10pt}\\
where $K$ denotes the Bayes factor and is the ratio of model evidences. The Bayes factor is usually more accessible than $R$ so it is usually used for model selection.\\
%
%

Bayesian regression has several positive and negative attributes. As a byproduct, Bayesian regression can perform model selection in a principled way that allows one to incorporate their prior knowledge into the selection process using BMT. Because Bayesian regression requires regressing probability distributions rather than just single model predictions, it can become intractable to calculate in general if the number of dimensions are large (as would standard regression if uncertainty is taken into account). Regularization in Bayesian regression is interpreted as coming from the uncertainty of the data and the uncertainty present in the prior parameters \cite{Bishop2006}, which we view as being a potential drawback. If one wants to change the regularization it would require changing either of these uncertainties, or both, ``artificially'' because one would be tuning their prior probabilities \emph{after} regression, which is a bit anti-Bayesian. Regularization can lead to an unnatural reduction of the posterior variances of the parameters for parametric models (a common problem with standard regression).\\

Further, we highlight some technical gaps found in Bayesian regression and model testing. Although almost all instances of Bayesian regression use data probability distributions that have infinite tails, truncated (or bounded) data probability density functions (pdfs) are realistic in practice too. We find that truncated data pdfs are potentially problematic for Bayesian regression if the model is deterministic. In the extreme case of completely certain data, Bayesian regression methods usually do not terminate because the Bayesian evidence is zero in (\ref{BayesianRegression}) since there are no possible combinations of parameter values that could exactly fit the data. This problem may also arise if the data uncertainties are bounded. In principle, standard regression methods can produce a solution regardless of the form of the data pdf. In what follows, we assume we are given a set of inputs $X$, a set of data points $Y = D$, and a set of model outputs $\hat{Y} = M(X; \vec{\alpha})$. All the sets are $n$-dimensional. In addition, we assume that the $n$ data points were collected through independent experiments. We give explicit examples of the likelihoods below (see Appendix \ref{appendixA}) given the model is deterministic:  \\

\noindent\underline{\textit{Infinite Tail Data Distributions}}\\
Data distributions with infinite tails result in likelihoods with infinite tails in (\ref{BayesianEvidence}). Some examples of infinite tail data distributions are Gaussian, Student-t, Laplace, canonical, and Poisson. For example, Gaussian distributed data (see Figure \ref{fig1a}) naturally has an infinite tailed likelihood function,\vspace{-10pt}\\
\begin{align}
\mathcal{L}(\vec{\alpha}) &= \displaystyle \frac{1}{\sqrt{(2\pi)^n|\Delta|}}e^{\textstyle-\frac{1}{2}\big(M(X;\vec{\alpha}) - D\big)^T\Delta^{-1}\big(M(X;\vec{\alpha}) - D\big)},\label{normalbmt}
\end{align}
where $\Delta$ is the covariance matrix. Since the likelihood has infinite tails, the predicted model response $M(x_j; \vec{\alpha})$ has probabilistic flexibility around its corresponding data point $D_j$ because it is uncertain. Even far from $D$, Bayesian regression is capable of estimating the posterior probability distributions of the model parameters in question as they are nonzero.\\

\noindent\underline{\textit{Truncated Tail Data Distributions}}\\
Data distributions with truncated tails naturally lead to truncated likelihoods in (\ref{BayesianEvidence}). For example, if the uncertain data is bounded to a region and is uniformly distributed, i.e. $D_j \sim \mathcal{U}(a_j,b_j)$ (see Figure \ref{fig1b}),
then the likelihood function is,\vspace{-15pt}\\
\begin{align}
\mathcal{L}(\vec{\alpha}) &= \displaystyle \prod_{j=1}^n\:\frac{\Theta\big(a_j\leq M(x_j;\vec{\alpha}) \leq b_j\big)}{b_j - a_j},\label{uniformbmt}
\end{align}
where $\Theta(\cdot)$ is the indicator function. In other words, for the likelihood $\mathcal{L}(\vec{\alpha})$ to be nonzero, the predicted model response $M(x_j;\vec{\alpha})$ at $x_j$ must lie within the interval $\big[a_j, \:b_j\big]$ for all $j$ simultaneously. The function space defined by the model and uncertain parameters is constrained by the data. This can make the probability of estimating a regressed posterior probability distribution of the model parameters very small, and in some cases impossible, because the likelihood may evaluate to zero for almost all combinations of $\vec{\alpha}$. \\

This point is exaggerated if the data is completely certain or deterministic, because the likelihood function becomes\vspace{-10pt}\\
\begin{align}
\mathcal{L}(\vec{\alpha}) &= \displaystyle \delta\big(M(X;\vec{\alpha}) - D\big),\label{certainbmt}
\end{align}
where $\delta(\cdot)$ is the Dirac delta function. In this case, the model output and observed data only agree if their values are exactly equal, i.e. $ M(X;\vec{\alpha})\rightarrow D$ for all $n$ points, which in most cases, is only possible if we overfit the data or the model is perfect. Thus, Bayesian regression will usually fail in this case, or if it succeeds, it only produces singular posterior distributions of the model parameters (i.e. $\sigma_{\vec{\alpha}}=0$). When Bayesian regression fails, the Bayesian evidence is zero, which, although correct (the model does not support/fit the data), may not be the most useful type of answer for the modeler. It seems reasonable that a modeler would want both the benefits of Bayesian and standard regression simultaneously.

\subsection{BVM Model Testing\label{section2.3}}

We present the Bayesian Validation Metric (BVM) proposed in \cite{Vanslette2019}. 
The BVM represents model to data validation in a general way using the probability of agreement, 
\begin{align}
p(A|M,D,B) &= \displaystyle \int_{\hat{z},z} p(A|\hat{z},M,z,D,B)\cdot\rho(\hat{z},z|M,D)\,d\hat{z}\,dz\notag\vspace{5pt}\\
&= \displaystyle \int_{\hat{z},z} \rho(\hat{z}|M,D)\cdot\Theta\big(B(\hat{z},z)\big)\cdot\rho(z|D)\,d\hat{z}\,dz,\label{bvm}
\end{align}
where $\hat{z}$ and $z$ are the model and data comparison quantities, respectively. The ``agreement kernel" $p(A|\hat{z},M,z,D) = \Theta\big(B(\hat{z},z)\big)$ is the indicator function of a user-defined Boolean function, $B(\hat{z},z)$, that defines the context of what is meant by ``model to data agreement" by being true when $(\hat{z},z)$ agree or false otherwise.  For simplicity, we will assume $\hat{z} \rightarrow \hat{y}$ and $z \rightarrow y$ are the model output and observed data respectively.\\ 

The BVM model testing framework was shown to generalize BMT where the probability of agreement plays the role of the evidence,\vspace{-15pt}\\
\begin{align}
\mathcal{Z}(B) = p(A|M,D,B) = \displaystyle\int_{\vec{\alpha}}\underbrace{\displaystyle p(A|\vec{\alpha},M,D,B)}_{\displaystyle \mathcal{L}(\vec{\alpha},B)}\cdot\underbrace{\vphantom{\displaystyle p(A|\vec{\alpha},M,D,B)}\rho(\vec{\alpha}|M)\,d\vec{\alpha}}_{\displaystyle \mathcal{\pi(\vec{\alpha})}\,d\vec{\alpha}}.\label{BVMEvidence}
\end{align}
where $\mathcal{Z}(B)$ and $\mathcal{L}(\vec{\alpha},B)$ are the BVM evidence and likelihood, respectively, that have been modified by a user's definition of model-data agreement $B$. Analogous to the Bayesian model testing framework, we can perform BVM model testing between two models $M_1$ and $M_2$ using the probability of agreement defined above as follows\vspace{12pt}\\ 
\centerline{$\displaystyle R(B)\equiv\frac{p(M_1|A,D,B)}{p(M_2|A,D,B)} = \frac{p(A|M_1,D,B)\:p(M_1|D,B)}{p(A|M_2,D,B)\:p(M_2|D,B)} = \frac{\mathcal{Z}_1(B)}{\mathcal{Z}_2(B)}\frac{p(M_1|D,B)}{p(M_2|D,B)},$}\vspace{12pt}\\
where $p(M_1|D,B)/p(M_2|D,B)$ is the ratio of prior probabilities of $M_1$ and $M_2$, which can often be set to unity, i.e. $p(M_1|D,B)/p(M_2|D,B) = 1$. In this case, we get\vspace{-10pt}\\
\begin{align}
R(B) \rightarrow \displaystyle \frac{p(A|M_1,D,B)}{p(A|M_2,D,B)} = \frac{\mathcal{Z}_1(B)}{\mathcal{Z}_2(B)} = K(B)\label{bvmratio}
\end{align}
where $R(B)$ denotes the BVM ratio and $K(B)$ denotes the BVM factor, which is analogous to Bayes factor.

\subsection{Example: Improved Reliability Metric\label{reliability}}
The reliability metric discussed in \cite{Rebba} is defined as the probability that the mean of the model prediction is within a tolerance $\epsilon$ of the mean of the data. This metric was later expanded in \cite{Sankararaman2} to consider tolerances between each of the model and data point pairs, $|y_j-\hat{y}_j|\leq \epsilon_j$ for $j=1,...,n$, rather than comparing their means.\\

Consider a set of inputs $X$, a set of model outputs $\hat{Y}$ and a set of observed data points $Y$. Assume that all the sets are $n$-dimensional and that the data were collected through independent experiments. The improved reliability metric $r_i$ is,\vspace{-10pt}\\
\begin{align}
r_i = \displaystyle\int_{Y}\rho(Y|D)\bigg(\int_{Y-\epsilon}^{Y+\epsilon}\rho(\hat{Y}|M)\,d\hat{Y}\bigg)\,dY,
\end{align}
which can always be rewritten as,\vspace{-10pt}\\
\begin{align}
r_i = \int_{\hat{Y},Y}\rho(\hat{Y}|M)\cdot\Theta\Big(\big|\hat{Y} - Y\big| \leq \epsilon\Big)\cdot\rho(Y|D)\,d\hat{Y}dY.
\end{align}
This equation may be identified as a special case of the BVM (\ref{bvm}) when,\vspace{-10pt}\\
\begin{align}
\Theta(B(\hat{z},z)) \rightarrow\Theta\big(B(\hat{Y},Y)\big)= \Theta\Big(\big|\hat{Y} - Y\big| \leq \epsilon\Big) = \displaystyle\prod_{j=1}^n \Theta\Big(\big|\hat{y}_j - y_j\big| \displaystyle\leq \epsilon_j\Big),\label{epsbool}
\end{align}
and where $\hat{z} = \hat{Y}$, $z = Y$ \cite{Vanslette2019}. Thus, this agreement kernel is based on the $\epsilon$-Boolean, as we call it in later sections.
From (\ref{BVMEvidence}) (see Appendix \ref{appendixB} for details), the BVM in this case is,\vspace{-10pt}\\
\begin{align}
\mathcal{Z}(B) = p(A|M,D,B)= \displaystyle\int_{\vec{\alpha}}\bigg( \int_{Y}\Theta\Big(\big|M(X;\vec{\alpha}) - Y\big| \leq \epsilon\Big)\cdot\rho(Y|D)\,dY\bigg)\cdot\rho(\vec{\alpha}|M)\,d\vec{\alpha}.\label{reliabilityevidence}
\end{align}
Thus, the choice of the Boolean agreement function results in representing the different validation metrics in terms of the BVM \cite{Vanslette2019}. The $\epsilon$-Boolean participates in several of our BVM regression examples in the following sections and thus reliability is automatically regressed into our model solutions through the improved reliability metric.

\section{Generalized Bayesian Regression via the BVM\label{bvmregression}}

This section introduces BVM regression, which generalizes Bayesian and standard regression. This method has the ability to produce posterior parameter distributions and predictive envelopes for any data distribution, include prior knowledge about model parameters (if there is any), and regularize parameter solutions in a way that parameter uncertainty increases rather than decreases.

BVM regression consists of learning the posterior of a set of parameters $\vec{\alpha}$, given the agreement $A$ and the Boolean function $B$, from the prior via Bayes' Rule,\vspace{-10pt}\\
\begin{align}
\rho(\vec{\alpha}|M)\stackrel{*}{\longrightarrow}\mathcal{P}(\vec{\alpha}|A)\equiv\rho(\vec{\alpha}|A,M,D,B) = \displaystyle \frac{p(A|\vec{\alpha},M,D,B)\,\rho(\vec{\alpha}|M)}{p(A|M,D,B)}\label{BVMRegression}.
\end{align}
After learning the posterior distribution of the model parameters, we can evaluate the predictive distribution defined by:\vspace{-10pt}\\ 
\begin{align}
p(\hat{y}|A,M,B) = \displaystyle\int_{\vec{\alpha}} p(\hat{y}|\vec{\alpha},A,M,B)\cdot \rho(\vec{\alpha}|A,M,D,B) \:d\vec{\alpha}.\label{BVMPredict}
\end{align}
Performing BVM regression requires evaluating the BVM probability of agreement. At the beginning of Appendix \ref{appendixB}, we give a derivation showing that (\ref{BVMEvidence}) can be written as,\vspace{-10pt}\\
\small
\begin{align}
\mathcal{Z}(B) = p(A|M,D,B)&= \displaystyle\int_{\vec{\alpha}}\bigg( \int_{Y}\Theta\Big(B\big(M(X;\vec{\alpha}),Y\big)\Big)\cdot\rho(Y|D)\,dY\bigg)\cdot\rho(\vec{\alpha}|M)\,d\vec{\alpha},\label{BVMEvidence2}
\end{align}
\normalsize
which is analogous to (\ref{BayesianEvidence}) in form,\footnote{In terms of the BVM, the Bayesian evidence in BMT, i.e. Equation (\ref{BayesianEvidence}), may be interpreted as the probability that the uncertain data and model output are exactly equal, i.e. $\rho(D|M) \equiv \rho(\hat{Y} \equiv Y|M,D) \equiv \rho(A|M,D)$ which is Equation (\ref{appceq1}) derived in Appendix \ref{appendixA}.} and where the comparison values are $\hat{z}=\hat{Y}=M(X;\vec{\alpha})$ and $z=Y$ (we assume we are dealing with a set of inputs, model outputs (i.e. a set of $n$ points on a model curve), and $n$ observed data points, where all the data points were collected through independent experiments). \\

BVM regression can reproduce Bayesian regression, standard regression, and likelihood-based methods as special cases. 
When the data and model outputs must be exactly equal to agree with one another \big(i.e. $\delta(\hat{Y} - Y)$\big), the BVM produces BMT as a special case and the regression solutions are given in Appendix \ref{appendixA}. Typical likelihood-based methods follow from the same ``exactly equal'' definition of model-data agreement. We find that the Boolean function 
$B_{S.R}(\hat{Y}(\vec{\alpha}^*),Y)$ that reproduces standard regression is defined to be true iff $\vec{\alpha}^* = \underset{\vec{\alpha}}{\mathrm{argmin}}\: \mathcal{E}\big(M(X;\vec{\alpha}),Y\big)$ for some objective function $\mathcal{E}(\cdot)$. This only gives nonsingular posterior parameter distributions and predictive model envelopes if the data is uncertain and/or if $\mathcal{E}\big(M(X;\vec{\alpha}),Y\big)$ does not have a unique global minimum. \\

If the objective function is convex, then we have a single minimum which results in one vector of parameters $\vec{\alpha}^*$ that makes $B_{S.R}(\hat{Y}(\vec{\alpha}^*),Y)$ true. However, when the cost function is non-convex, then multiple parameter vectors $\vec{\alpha}^*$ corresponding to different local minima, lead to a true $B_{S.R}(\hat{Y}(\vec{\alpha}^*),Y)$ and may be accepted due to the approximate nature of non-convex optimization methods. This results in multiple regressed solutions for the regression problem and approximates the posterior parameters' distribution $\rho(\vec{\alpha}|A)$ (analogous to the accepted parameter samples in MCMC simulation). Marginalizing leads to the predictive posterior model distribution $p(\hat{y}|A,M,B)$ as in (\ref{BVMPredict}). Finding the predictive model output average is analogous to the results obtained in the ensemble methods in machine learning \cite{ensemble}. 
Because the BVM can reproduce these special cases and generate new ones by extending, combining, and modulating Boolean agreement functions, BVM regression may be seen as a generalized regression method.\\ 
 
Due to the flexibility of the BVM framework, there are many possible definitions of agreement that the user can define. Table \ref{bool} below contains some of these definitions.
\begin{table}[H]
\footnotesize
\begin{center}
\begin{tabular}{ |r|c| }
\hline\xrowht[()]{5pt}
& Agreement Boolean Function \\
\hline\hline\xrowht[()]{10pt}
$\epsilon$--Boolean& True iff $\big|y_j - M(x_j;\vec{\alpha})\big| \displaystyle\leq \epsilon_j\;\forall \;j$\\\hline\xrowht[()]{10pt}
$(\gamma,\epsilon,\ell)$--Boolean& True iff $\big|y_j - M(x_j;\vec{\alpha})\big| \displaystyle\leq \ell \epsilon_j\;\forall \;j$ and $\frac{1}{n}\sum_j \Theta\Big(\big|y_j - M(x_j;\vec{\alpha})\big| \displaystyle\leq \epsilon_j\Big) \geq \gamma\%$\\
\hline\xrowht[()]{10pt}
$\langle\epsilon\rangle$--Boolean& True iff $ \frac{1}{n}\sum_j\big|y_j - M(x_j;\vec{\alpha})\big| \displaystyle\leq \langle\epsilon\rangle$\\
\hline\xrowht[()]{10pt}
$(\langle\epsilon\rangle, \hat{\alpha})$--Boolean &  True iff $ \frac{1}{n}\sum_j\big|y_j - M(x_j;\vec{\alpha})\big| \displaystyle\leq \langle\epsilon\rangle$ and $0.91 \leq \frac{1}{n}\sum_j \Theta(y_j\in[-c_{\hat{\alpha}}, c_{\hat{\alpha}}]_j) \leq 0.99$\\
\hline
\end{tabular}
\vspace*{5pt}
\caption{Some examples of agreement Boolean functions.\label{bool}}
\end{center}
\end{table}
\noindent\vspace{-30pt}\\
To address the concerns we raised about Bayesian and standard regression depicted in Figure \ref{fig1}, consider using the $\epsilon-$Boolean with the agreement kernel,\vspace{8pt}\\
\indent\centerline{$\Theta\Big(B\big(M(x_j;\vec{\alpha}),y_j\big)\Big)$= \begin{math}
    \begin{dcases}
        1, & \text{if } \big|y_j - M(x_j;\vec{\alpha})\big| \displaystyle\leq \epsilon_j \\[\smallskipamount]
        0, & \text{otherwise}\\
    \end{dcases}
\end{math}}
\noindent\vspace{0pt}\\for all $j = 1,\hdots,n$, where $\epsilon_j$ may be adjusted and tuned to impose more or less strict agreement conditions which may be used by the modeler to enforce reliability in some region or to be more tolerant of training errors at instance $x_j$. For simplicity, We assume that $\epsilon_j = \epsilon$ for all $j$. Note that this is (\ref{epsbool}) in Section \ref{reliability}. In other words, we use the special case of the BVM, the improved reliability metric with evidence derived in (\ref{reliabilityevidence}), to derive our theoretical solutions.\footnote{Note that by choosing to adopt a different agreement kernel (or Boolean function) as in Section \ref{predex}, we generalize the results derived above; this is the power of the BVM.} Utilizing this BVM definition allows us to solve the truncated tail data distributions problem in Bayesian regression in a simple way -- details in Appendix \ref{appendixB}.\footnote{A complete derivation for the infinite tail Gaussian data distribution is given in Appendix \ref{appendixB1}.}\vspace{5pt}\\ 
\noindent\underline{\textit{Truncated Tail Data Distributions Solution Summary}}\\
Let the data be known to have the truncated pdf $y_j \sim \mathcal{U}(a_j,b_j),$ for $ j = 1,\hdots,n$. By using the $\epsilon$-Boolean, we introduce leniency into the regression in that it no longer needs to exactly pass through all intervals $[a_j,b_j]$ simultaneously to count as a ``fit''. This produces likelihood functions such as,\vspace{-10pt}\\
\begin{align}
\mathcal{L}(\vec{\alpha},B) &= \prod_{j=1}^n\:\frac{u_j - l_j}{b_j - a_j},\label{uniformbvm}
\end{align}
where $l_j$ and $u_j$ are defined by the boundaries of the intersection of the data uncertainty and the model's tolerance $\epsilon$,\vspace{8pt}\\
\centerline{$\Big[l_j,\: u_j\Big] = \Big[M(x_j;\vec{\alpha}) - \epsilon,\: M(x_j;\vec{\alpha}) + \epsilon\Big] \cap \Big[a_j,\: b_j\Big] \qquad j=1,\hdots,n.$}\vspace{8pt}\\
\indent An illustration of how the BVM works with truncated data distributions is shown in Figure \ref{fig2} below. For example, at instance $x_2$, the interval $\big[l_2,\: u_2\big]$ is found by intersecting the intervals $\big[a_2,\: b_2\big]$ and $\big[M(x_2;\vec{\alpha}) - \epsilon,\: M(x_2;\vec{\alpha}) + \epsilon\big]$. Note that this applies to the instances $x_j$ for all $j$. In this case, the likelihood is nonzero, resulting in a nonzero evidence (\ref{BVMEvidence2}). Thus, given this agreement definition, the probability of finding a model given the truncated data is nonzero.
\begin{figure}[H]
\indent 
  \centerline{\includegraphics[width=8cm]{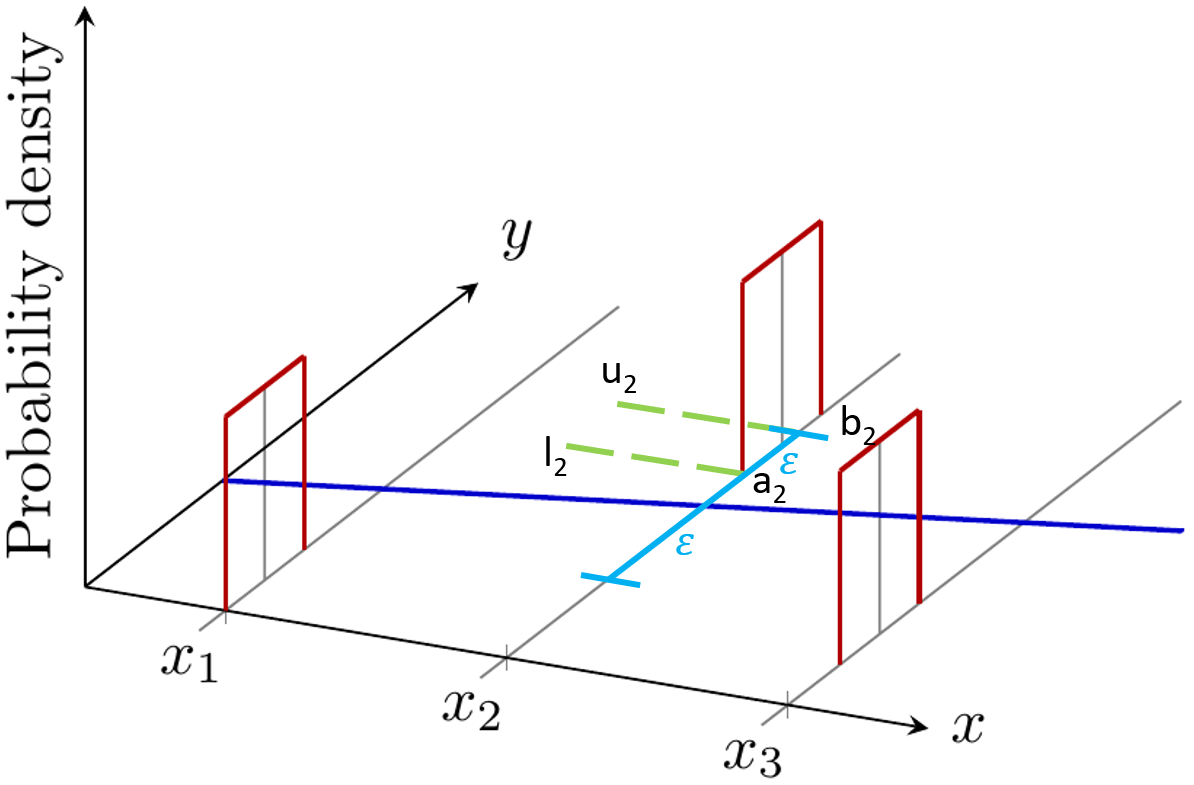}}
\caption{Truncated tail data distributions solution. Using BVM regression results in a nonzero probability of finding a model given the observed truncated data.\label{fig2}}
\end{figure}

\indent Now, if we consider the special case when the data is completely certain, deterministic, i.e. $Y = D$, then the likelihood function is\vspace{-15pt}\\
\begin{align}
\mathcal{L}(\vec{\alpha},B) &= \Theta\Big(B\big(M(X;\vec{\alpha}),D\big)\Big),\label{certainbvm}
\end{align} 
which can be seen as a relaxed general form of the delta function adopted in the Bayesian model testing (where $\epsilon = 0$), which implies that the model output must be within $\epsilon$ from the observed measurements in order for them to agree.  An analogous $\epsilon$-Boolean solution exists for standard regression methods which leads to nonsingular parameter distributions whether the regression is regularized or not.\vspace{5pt}\\ 
\noindent\underline{\textit{Tolerant agreement as a new kind of regularization}}\\
\indent The purpose of regularization is to better represent one's expectations of unobserved data using the chosen model or model class. Using BVM regression and nonzero agreement tolerances (e.g. $\epsilon>0$ in the $\epsilon$-Boolean), we can broaden the model's prediction envelope to better represent our expectations of the data.  Increasing agreement tolerances naturally increases the posterior variance of the parameters, which differs from standard regularization methods and can be used to avoid conceptual issues of interpreting regularized physical parameters. It should also be noted that this is done without changing the prior distributions of the parameters nor the given probability distributions of the data. This becomes a useful feature in our first example.

\section{Implementation and Examples\label{simulation}}


\subsection{Implementation Procedure\label{section3.2}}

Like the Bayesian evidence, the BVM evidence is computationally expensive to calculate when one has many model parameters to learn. Several approaches were adopted to solve this problem. Markov Chain Monte Carlo (MCMC) is a computational technique used for Bayesian methods that has been widely studied and improved \cite{bayesmcmc, Hasting, Metropolis, Neal, Marzouk, mcmcacc} as it is considered an indispensable tool for Bayesian inference. Other techniques include the Nested Sampling method \cite{FH08,Skilling2004} and the MultiNest algorithm \cite{FH09}.\\ 

We will approximate the BVM evidence and generate the posterior model parameter distributions (for the purpose of generating model's predictive envelopes) using MCMC (with proposal or candidate samples being accepted/rejected according to Metropolis-Hastings criterion \cite{mhalgo,Hasting}). The MCMC algorithm takes the following main inputs: the likelihood function $\mathcal{L}$ and the prior distribution $\pi(\vec{\alpha})$ of the model parameters, a model $M$ and a set of input/output data points $\{X, Y\}$. The Bayesian terms $(\mathcal{Z},\mathcal{L})$ have analogous BVM terms $(\mathcal{Z}(B),\mathcal{L}(B))$ and it is therefore straightforward to extend the MCMC algorithm to BVM calculations to obtain posterior parameter distributions. These distributions are used to generate a model's predictive envelope. Note that the MCMC algorithm takes some additional inputs (e.g. the number of iterations, burn-in, etc.) that are usually optional or chosen according to the user's preference and the problem setting. \\

As we outlined above, our generalized Bayesian regression method is simple and easy to implement. Algorithm \ref{algo1} summarizes the implementation procedure that we will adopt later in the illustrative numerical case study and examples. 

\begin{algorithm}[H]
\caption{Implementation Procedure\label{algo1}}
\begin{enumerate}[leftmargin=0.8cm]
\item Prepare the data  $\{X, Y\}$ to be used for regression/calibration.
\item Determine the model $M$ (to fit the data) and its parameters' prior distributions $\pi(\vec{\alpha})$.
\item Choose the comparison quantities (usually the model ouptut and the observed data) and an appropriate Boolean agreement function $B$ (it is recommended to include the reliability metric inside the agreement function).
\item Derive the likelihood function $\mathcal{L}(B)$ (after specifying the type of data uncertainty) and determine a method to compute it, either analytically or numerically.
\item Compute the BVM evidence $\mathcal{Z}(B) \longleftarrow \texttt{MCMC}\big(\{X, Y\},\mathcal{L}(B),M,\pi(\vec{\alpha}),etc.\big)$. This will produce the parameters' posterior distributions $\mathcal{P}(\vec{\alpha})$.
\end{enumerate}
\end{algorithm}

\subsection{BVM Regression Examples\label{section3.3}}
\subsubsection{Numerical Example 1: Bacterial Growth Model\label{bacterialex}}
We consider the case study investigated in \cite{Brown2002} using a bacterial growth model. The data is obtained by operating a continuous flow biological reactor at steady-state conditions. The observations are as follows:
\begin{table}[H]
\label{Table1}
\begin{center}
 \begin{tabular}{ l c c c c c c c } 
 \toprule
$\boldsymbol{x}$ \textbf{(mg/L COD)}\:\: & 28 & 55 & 83 & 110 & 138 & 225 & 375 \\ [0.7ex] 
$\boldsymbol{y}$ \textbf{(1/h)}\:\: & 0.053 & 0.060 & 0.112 & 0.105 & 0.099 & 0.122 & 0.125 \\[0.9ex]
 \bottomrule
\end{tabular}
\vspace*{5pt}
\caption{The observations we aim to fit.\label{t2}}
\end{center}
\end{table}
\noindent\vspace{-30pt}\\
\noindent where $y$ is the growth rate at substrate concentration $x$. We replicate the results found in \cite{Brown2002} using the nonlinear Monod model to fit the data, i.e,\vspace{-10pt}\\
\begin{align}
    \hat{y} = M(x;\vec{\alpha}) = \displaystyle\frac{\alpha_1x}{\alpha_2 + x}\label{monod}
\end{align}
where $\alpha_1$ is the maximum growth rate (h$^{-1}$: per hour), and $\alpha_2$ is the saturation constant (mg/L COD: the Chemical Oxygen Demand, measured in milligrams per liter).\\ 

Following the procedure outlined in Section \ref{section3.2} is straightforward. We first prepare the data: from Table \ref{t2}, we have $\{X, Y\} = \{\boldsymbol{x}, \boldsymbol{y}\}$. We then determine the model $M$ which, in this case, is described by Equation (\ref{monod}). To determine the prior distribution of the parameters $\pi(\alpha_1)$ and $\pi(\alpha_2)$ (if no prior knowledge about the parameters is known), a good practice is to assume that the prior distribution is Gaussian around a point close to the solution resulting from applying least squares (i.e. minimizing the sum of the squares of the residuals), with some standard deviation. In our case, the least squares solution is $\alpha_1 = 0.14542$ and $\alpha_2 = 49.053$, and thus we assume $\pi(\alpha_1) = \mathcal{N}(0.17, \sigma^2_{\alpha_1})$ and $\pi(\alpha_2) = \mathcal{N}(47.5, \sigma^2_{\alpha_2})$, where $\sigma_{\alpha_1}$ and $\sigma_{\alpha_2}$ are determined by the user (here, we use $\sigma_{\alpha_1} = 0.025$ and $\sigma_{\alpha_2} = 3$). We then choose the comparison quantities to be the sets of model outputs and observed data, i.e. $\hat{z} \rightarrow \hat{Y}$ and $z \rightarrow Y$, and use the $\epsilon-$Boolean agreement function (i.e. the improved reliability metric is included in our agreement definition). Regarding the likelihood function, we experiment with the different types of data uncertainty that were discussed earlier. In other words, we use the likelihoods derived in (\ref{uniformbvm}) and (\ref{certainbvm}) corresponding to the different types of data distributions discussed above, i.e. bounded or truncated uniform data distributions, and completely certain observation points (we also consider normal or Gaussian data distributions with infinite tails). Finally, we run MCMC and find that BVM regression is capable of constructing posterior inferences for the model parameters for each type of data distributions (i.e. for each likelihood function that we use), unlike Bayesian model testing and standard regression techniques that fail at this task for truncated and completely certain data, as discussed before. We summarize the results in Table \ref{Table2} below.
\begin{table}[H]

\begin{center}
 \begin{tabular}{lccc} 
 \toprule
 Data Distribution & BVM regression & Standard regression & Bayesian regression \\ [0.5ex] 
 \midrule
Infinite Tail & \ding{51} & \ding{51} & \ding{51} \\ 
 Truncated Tail & \ding{51} & \ding{51} & \ding{55} \\
 Completely Certain & \ding{51} & \ding{55} & \ding{55} \\
 \bottomrule
\end{tabular}
\vspace*{5pt}
\caption{High probability of the model producing posterior parameter distributions and predictive envelopes for different types of data distributions using the three approaches. BVM regression is capable of producing posterior distributions of the model parameters for any type of data distributions.\label{Table2}}
\end{center}
\end{table}
\noindent\vspace{-35pt}\\
\indent Using the BVM regressed parameters' distributions, we can make predictions of $y$ for new values of $x$, i.e., $p(\hat{y}|A,M,B)$ from (\ref{BVMPredict}). In addition, instead of just computing a point estimate of the fit, we also study the predictive posterior distribution of the model (also called the predictive envelope or confidence band) that results from the posterior distributions of the parameters. As an illustration of the predictive posterior distribution of our BVM regressed model, we plot the predictive envelopes of the nonlinear Monod model described in (\ref{monod}), along with the joint posterior distribution of the parameters $\alpha_1$ and $\alpha_2$, while treating the data as completely certain and using the $\epsilon$-Boolean function with a tolerance $\epsilon = 0.03$.\vspace{-15pt}\\

\begin{figure} [H]
\centering
\begin{subfigure}[b]{8.4cm}
\includegraphics[width=8.4cm]{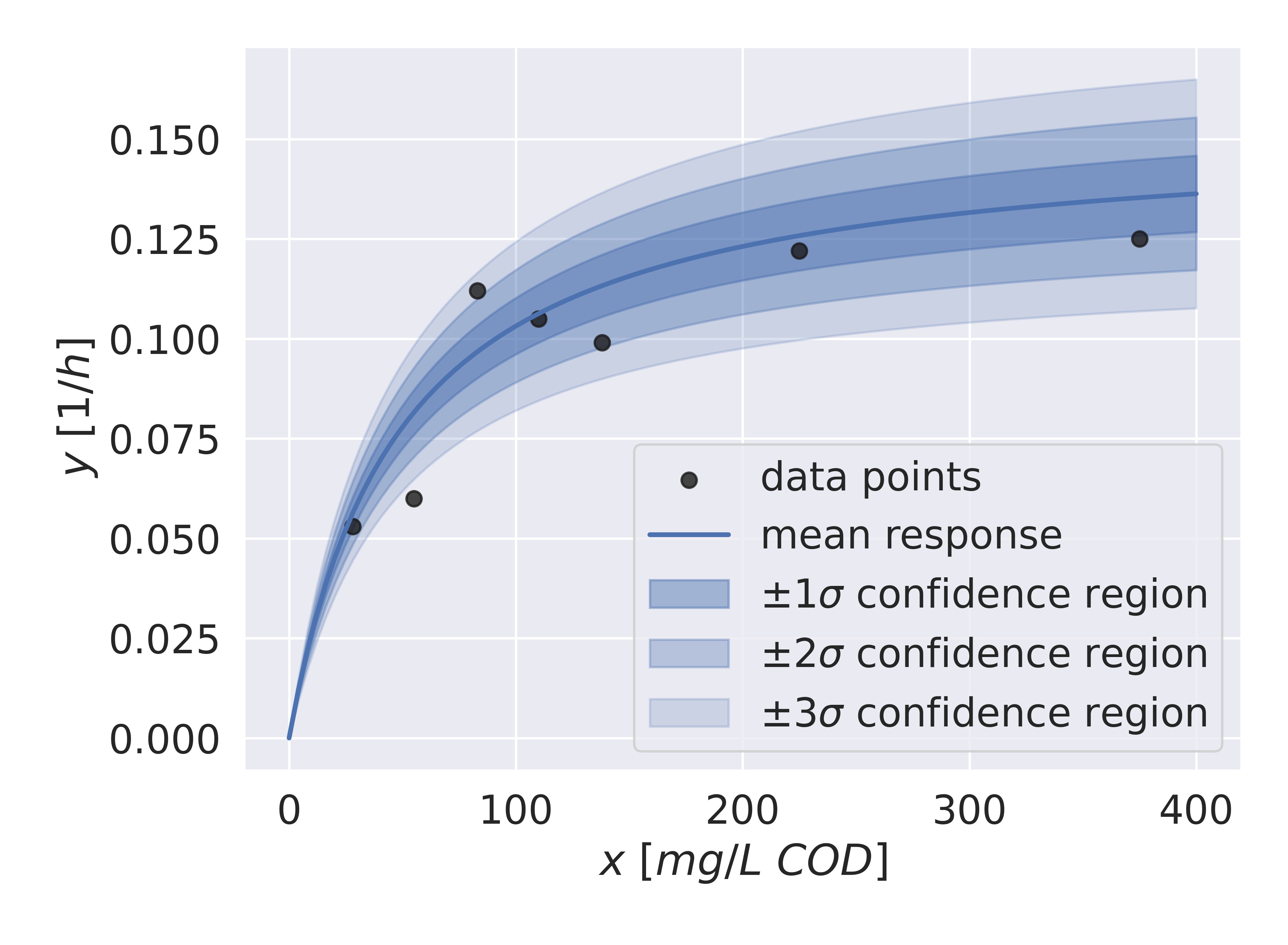}
\end{subfigure}
\begin{subfigure}[b]{7cm}
\includegraphics[width=7cm]{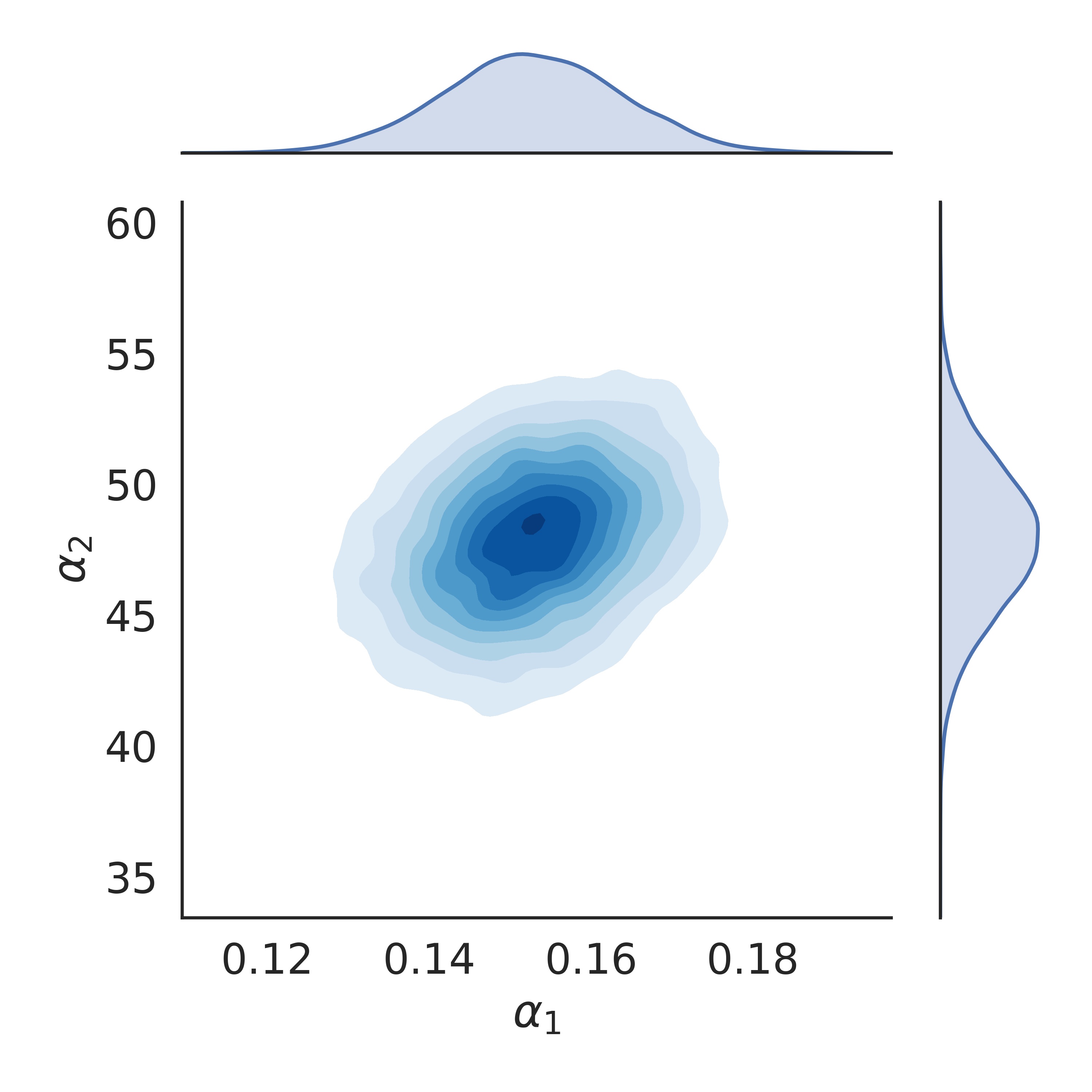}
\end{subfigure}
\caption{Predictive envelopes of the model and parameters' joint posterior distribution in the absence of data uncertainty using the BVM. As tabulated in Table \ref{Table2}, Bayesian regression fails to produce a candidate model solution as the data is completely certain and standard regression produces a single deterministic solution with no model uncertainty.\label{jointplot}}
\end{figure}

The blue curve shows the predicted response, which is the model fit calculated using the mean values of the parameters $\alpha_1$ and $\alpha_2$ in the chain. The blue shaded areas correspond to 68\%, 95\%, and 99.7\% predictive posterior regions (by computing the model fit for a randomly selected subset of the chain). In other words, the blue regions span 1, 2, and 3 standard deviations on either side of the mean response, respectively. We will leave the interpretation of the predictive envelopes for our compound Boolean agreement function example in Section \ref{predex}.\\

By examining the model parameters' joint posterior distribution in Figure \ref{jointplot}, one can suspect that the value of the tolerance $\epsilon$ chosen affects the shape of the model parameters' distributions and thus the predictive envelope (or confidence regions). A smaller tolerance implies stricter agreement conditions between the model outputs and the observed data, which results in less uncertainty in the predictive posterior distributions of the model parameters and a narrower envelope. On the other hand, a larger tolerance implies more flexible agreement conditions, and results in more uncertainty in the predictive distributions, a wider envelope and a less predictive power. Thus, increasing $\epsilon$ can always result in finding a model given the data. To avoid getting very wide envelopes relative to the spread of the data, we start with a very small $\epsilon$ when running the MCMC simulation. We then keep increasing $\epsilon$ until the MCMC algorithm starts achieving a reasonably small acceptance rate for the new candidates in the chain.\\

\begin{figure} [H]
\centering
\includegraphics[width=16.6cm]{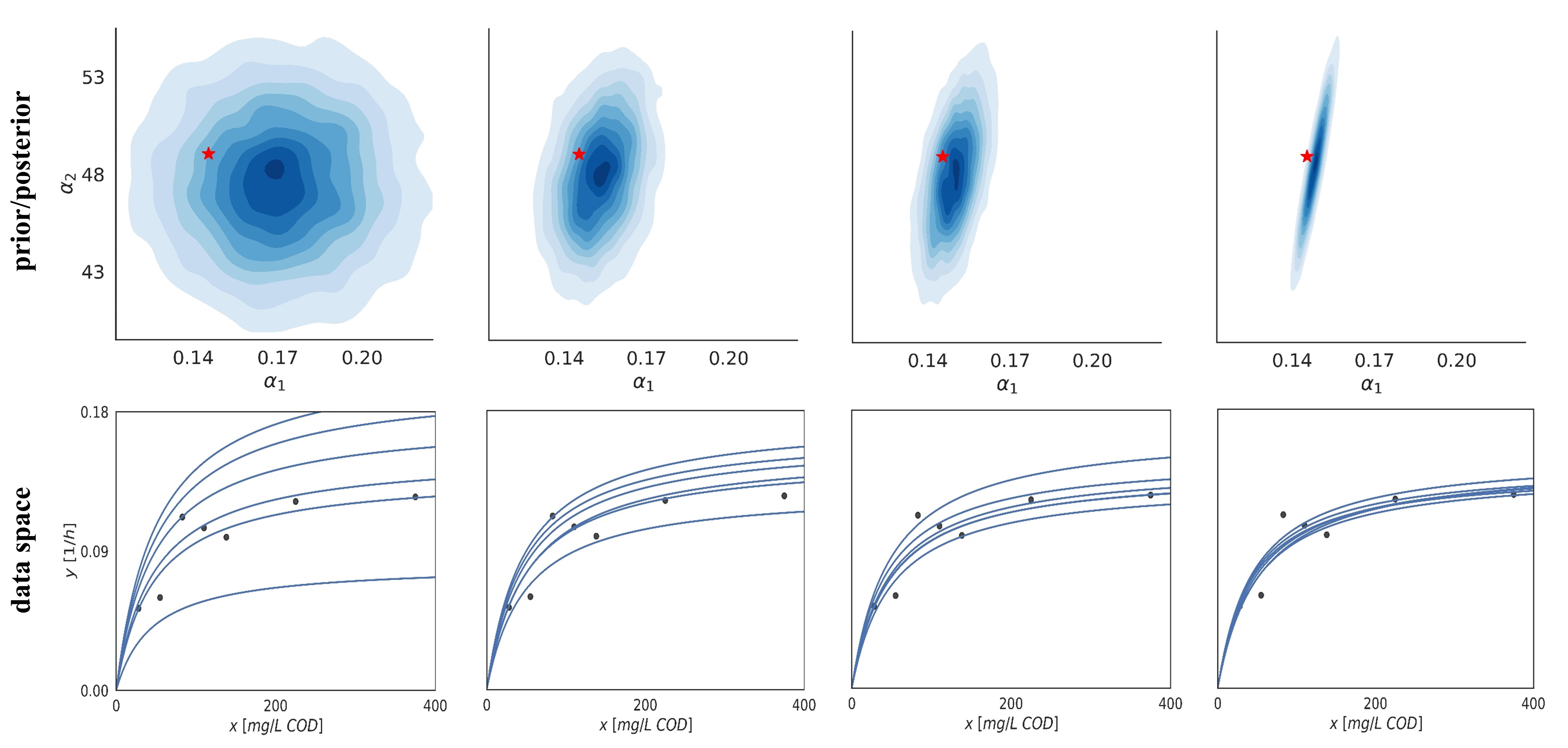}
\caption{Illustration of BVM Learning for the Monod model for decreasing values of $\epsilon$. In the first row is the prior/posterior parameter distribution in $(\alpha_1,\alpha_2)$ space. The data points are shown by a black circle in the second row. The first column corresponds to the situation before any data point is observed and shows a plot of the prior distribution in $(\alpha_1,\alpha_2)$ space together with six samples of the model response $M(x; \vec{\alpha})$ (blue lines) in which the values of $\alpha_1$ and $\alpha_2$ are randomly drawn from the prior. In the second, third and fourth columns, we see the situation after running BVM regression using MCMC, with a tolerance $\epsilon = 0.03$, $\epsilon = 0.025$ and $\epsilon = 0.02$, respectively. The posterior has now been influenced by the agreement tolerance $\epsilon$, this gives a relatively compact posterior distribution. Samples from this posterior distribution lead to the functions shown in blue in the second row.
\label{post}}
\end{figure}

Since this model has just two adaptive parameters, namely $\alpha_1$ and $\alpha_2$, we can plot the prior and posterior distributions directly in parameter space. We explore the dependence between the parameters' posterior distributions and the value of the tolerance $\epsilon$. Figure \ref{post} shows the results of BVM learning for the Monod model in (\ref{monod}) as the value of $\epsilon$ is decreases. For comparison, the optimal parameter values $\alpha_1 = 0.14542$ and $\alpha_2 = 49.053$ computed using standard regression are shown by a red star in the first row of Figure \ref{post}.
As Figure \ref{post} shows, the smaller the tolerance is, the narrower and sharper the posterior joint distribution of the parameters is, the closer the blue lines get to each other, and the lower the uncertainty is. This explains the shape of the predictive envelopes as was discussed before. Thus, by varying $\epsilon$, one can tune the posterior distribution of the model response to be more or less representative of the data. We will elaborate more on this in Section \ref{discussion}.
It is worth noting that for a very small value of $\epsilon$, the posterior joint distribution of the parameters converges toward the least squares solution, highlighted by a red star in figure \ref{post}. Also note that in our example, when $\epsilon$ goes below about $0.017$, no solution seems to be possible and hence the probability of finding a model given the observed data becomes zero (i.e. our method starts to behave like Bayesian regression). In this case, the analyst may choose to work with any tolerance beyond this threshold, depending on his specifications and agreement requirements.\\

Once we generate the posterior distributions of the model parameters and the predictive envelopes, we can measure and estimate the reliability and accuracy of our computational model using a validation metric (including the BVM). By doing so, we determine how accurate is our model representation of the real world.
\clearpage

\subsubsection{Numerical Example 2: Toy Model\label{predex}}
After showing how the BVM can be used to perform regression on any type of data distribution to generate posterior model parameters' distributions and predictive envelopes, we now focus on how the user can choose the Boolean function to define the model-data agreement.\\

In this example, we will use the compound Boolean function as presented in \cite{Vanslette2019}.\footnote{It should be noted that in \cite{Vanslette2019}, a model is simply validated according to this compound metric -- here we calibrate the model with respect to it instead.} In that case, the definition of agreement requires the model to pass an average square error threshold of $\langle \epsilon \rangle$ as well as a check for probabilistic model configuration. The latter states that $95\% \pm 4\%$ of the uncertain (data) observations should lie inside the model's $1 - \hat{\alpha} = 95\%$ confidence interval. Note that we impose the $\pm 4\%$ tolerance to prevent the scenario where all $100\%$ of the data points lie within an overly wide confidence interval, being marked as ``agreeing''. We denote this compound Boolean function by $B(\hat{Y},Y, \langle \epsilon \rangle, \hat{\alpha})$ and it is equal to, \vspace{-10pt}\\
\small
\begin{align}
B\bigg(\frac{1}{n}\sum_i |\hat{y}_i - y_i| \leq \langle\epsilon\rangle\bigg) \wedge B\bigg(0.91 \leq \frac{1}{n}\sum_i \Theta(y_i\in[-c_{\hat{\alpha}}, c_{\hat{\alpha}}]_i) \leq 0.99\bigg),\label{compbool}
\end{align}
\normalsize
where $n$ is the number of data points in the set $Y$, and $[-c_{\hat{\alpha}}, c_{\hat{\alpha}}]_i$ is the model's $95\%$ confidence interval at instance $x_i$ (this is the $(\langle\epsilon\rangle, \hat{\alpha})$--Boolean in Table \ref{bool}). Note that, although this compound Boolean seems to be complex, it is relatively easy to code and implement.\\

\noindent The BVM probability of agreement in this case can be expressed as, \vspace{-10pt}\\
\small
\begin{align}
\mathcal{Z}(B) = p(A|M,D,B,\langle\epsilon\rangle, \hat{\alpha})&= \displaystyle\int_{\vec{\alpha}}\underbrace{\bigg( \int_{Y}\Theta\Big(B\big(M(X;\vec{\alpha}),Y,\langle\epsilon\rangle, \hat{\alpha}\big)\Big)\cdot\rho(Y|D)\,dY\bigg)}_{\textstyle\mathcal{L}(\vec{\alpha},B)}\cdot\rho(\vec{\alpha}|M)\,d\vec{\alpha}\notag
\end{align}
\normalsize
Note that the likelihood $\mathcal{L}(\vec{\alpha},B)$ can be expressed as an expectation value over $\rho(Y|D)$,\vspace{-10pt}\\
\small
\begin{align}
\mathcal{L}(\vec{\alpha},B) = E\Big[\Theta\Big(B\big(M(X;\vec{\alpha}),Y,\langle\epsilon\rangle, \hat{\alpha}\big)\Big) \Big] \sim \frac{1}{K}\sum_{k=1}^K \Theta\Big(B\big(M(X;\vec{\alpha}),Y^{(k)},\langle\epsilon\rangle, \hat{\alpha}\big)\Big),\label{MCeq}
\end{align}
\normalsize
where $Y^{(k)}$ denotes the $k^{th}$ set of data points drawn randomly from the probability distribution of $Y$. This allows us to approximate the integral using a statistical method like Monte-Carlo (MC). In this example, we use MC with $K = 50$.\\

We implement the compound Boolean, $B(\hat{Y},Y, \langle \epsilon \rangle, \hat{\alpha})$, and show its ability to combine and quantify the average error as well as the probabilistic model representation of the uncertain data observations. We generated data using,\vspace{-10pt}\\
\begin{align}
y(x) = 1 + x e^{\textstyle -\cos{(10x)}} + \sin{(10x)} + \epsilon_{a}(x),\notag
\end{align}
where $\epsilon_a(x) \sim \mathcal{N}(0,0.4^2)$ for $x\in[0,1.5]$ and $\epsilon_a(x) \sim \mathcal{N}(0,0.6^2)$ for $x\in [1.5,3]$, which represents the aleatoric stochastic uncertainty due to the system's randomness. We also assume the presence of epistemic measurement uncertainty in the data \cite{Sankararaman} with an additional normal distribution $\mathcal{N}(0,0.5^2)$ about each data point.\\

\begin{figure} [H]
\centering
\begin{subfigure}{8.2cm}
\includegraphics[width=8.2cm]{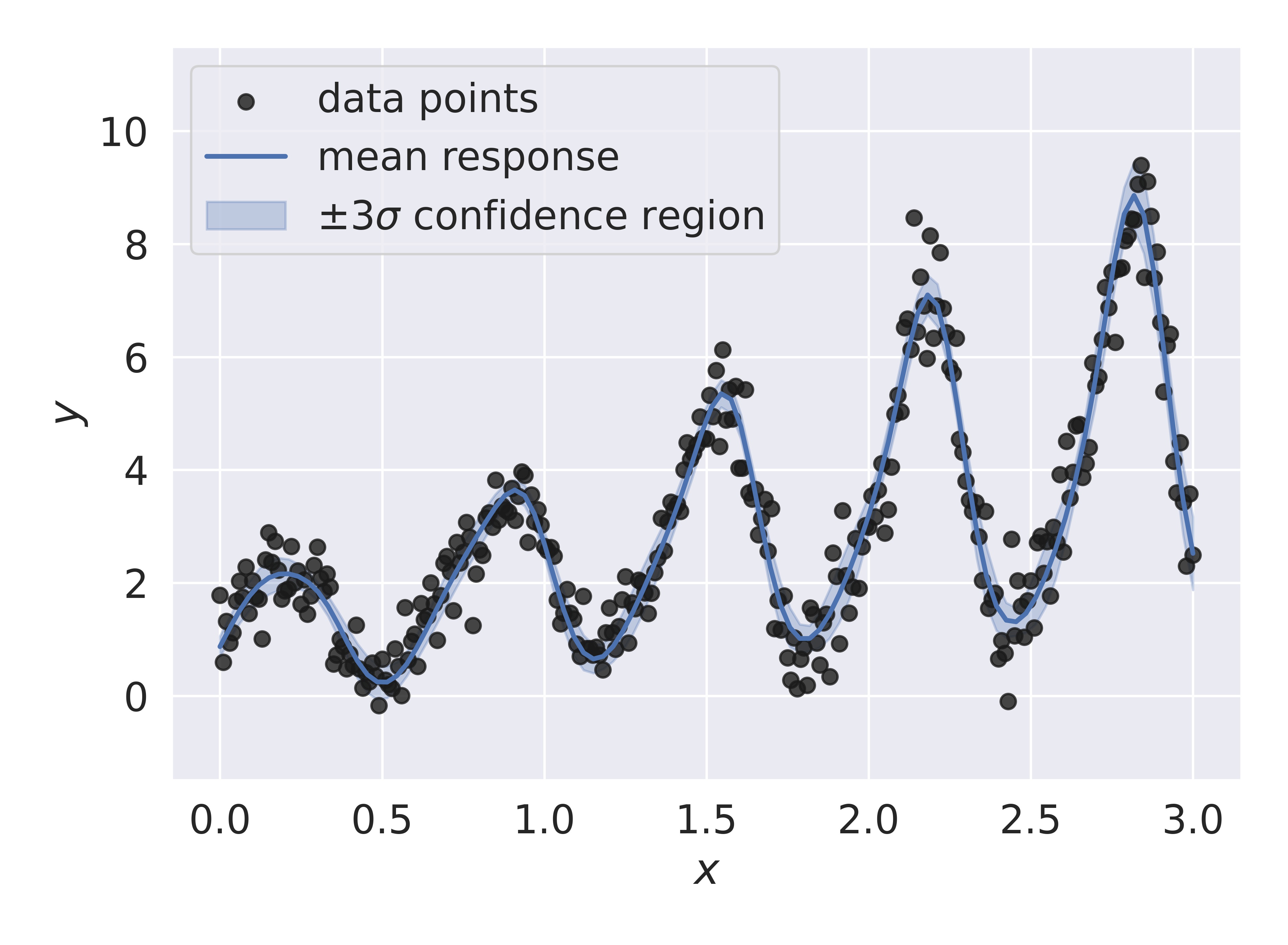}
\caption{Bayesian regression.\label{figg1a}}
\end{subfigure}
\hspace{-4pt}
\begin{subfigure}{8.2cm}
\includegraphics[width=8.2cm]{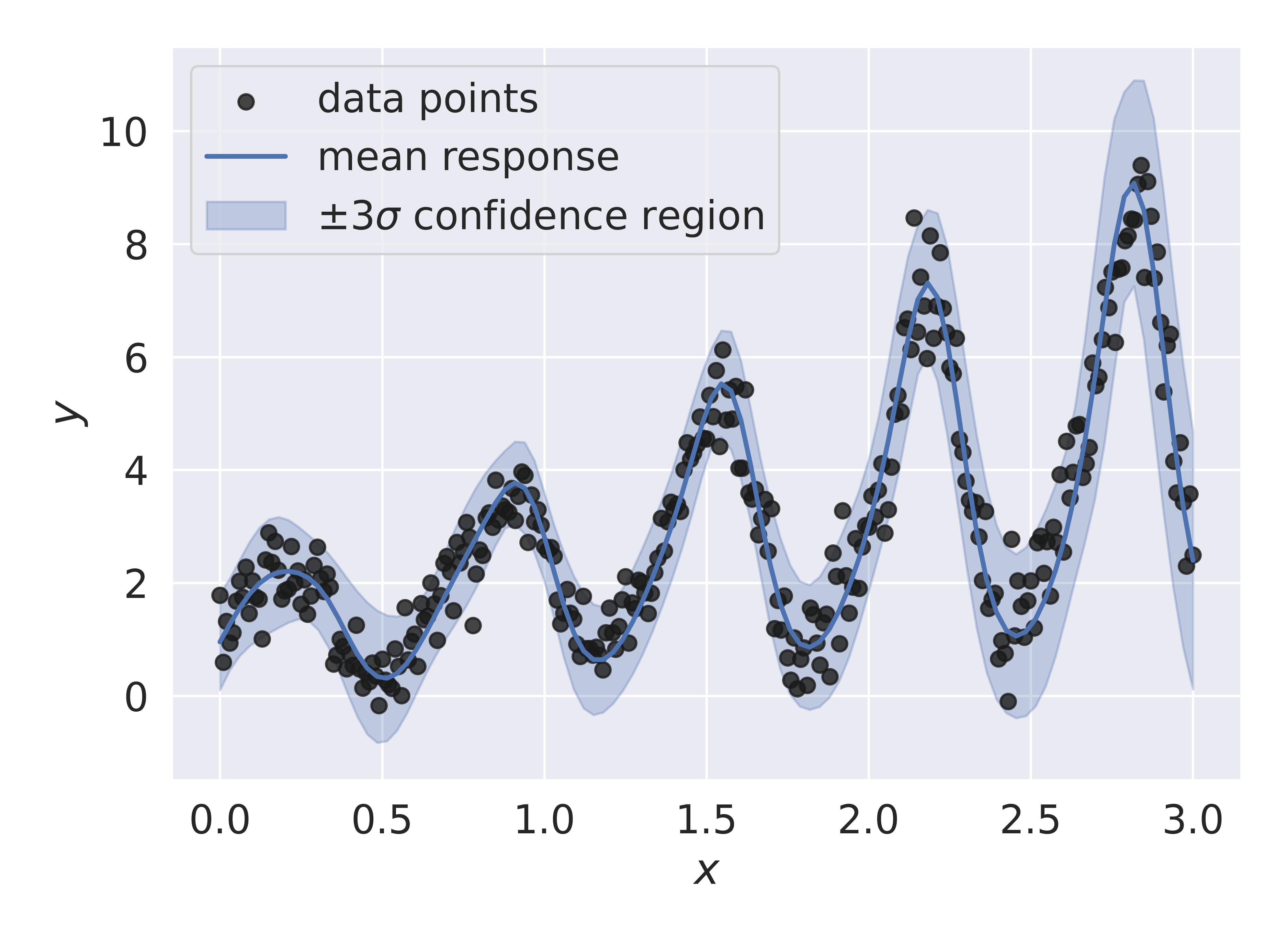}
\caption{BVM regression.\label{figg1b}}
\end{subfigure}
\caption{Comparison between Bayesian regression and BVM regression. (a) Bayesian regression under infinite tail (Gaussian) data distribution. Note that the $\pm 3\sigma$ confidence region is very narrow and standard regression method produces a nearly identical result. (b) BVM regression using the compound Boolean. In this case, the $\pm 3\sigma$ confidence region is much wider and represents the data more accurately. Note that this probabilistic model passes both agreement conditions imposed by the compound Boolean $B(\hat{Y},Y, \langle \epsilon \rangle, \hat{\alpha})$. Starting with a very small $\langle\epsilon\rangle$ in the MCMC simulation, we tune $\langle\epsilon\rangle$ by gradually increasing its value until both elements of the compound Boolean are naturally satisfied. \label{BVMvsBMT}}
\end{figure}

\noindent To solve this example, we consider the following deterministic non-linear model,\vspace{-10pt}\\
\begin{align}
\displaystyle\hat{y} = M(x;\vec{\alpha}) = \alpha_1 + \alpha_2x e^{\textstyle -\alpha_3\cos{(\alpha_4x)}} + \alpha_5\sin{(\alpha_6x)},\label{model2}
\end{align}
where $\vec{\alpha} = (\alpha_1,\alpha_2,\alpha_3,\alpha_4,\alpha_5,\alpha_6)$ is the vector of model parameters having normally distributed prior distributions with means $\mu_{\vec{\alpha}} = (0,0,0,9,0,9)$ and standard deviations $\sigma_{\vec{\alpha}} = (1,1,1,0.5,1,0.5)$.
We then conduct two experimental simulations while treating the data as normally distributed (i.e. having infinite tails). We first run the MCMC algorithm for 5000 iterations with $10\%$ burn-in using Bayesian regression and plot the results in Figure \ref{figg1a}. We then repeat the simulation by performing BVM regression instead (i.e. by following the procedure outlined in Section \ref{section3.2}), and using the approximate likelihood $\mathcal{L}(\vec{\alpha},B)$ from (\ref{MCeq}) with a threshold of $\langle\epsilon\rangle = 0.7$. The results are shown in Figure \ref{figg1b}.

The BVM regression framework offers new insights into the interpretation of the predictive envelopes of Bayesian and standard regression. It is clear in Figure \ref{figg1a} that the Bayesian and standard regression methods generate predictive envelopes that would not accurately predict new target points. Surprisingly, these envelopes actually quantify the uncertainties in the \emph{least square error solution} due to the presence of data uncertainty rather than a measure of predictive uncertainty. By choosing an appropriate model-data agreement function (as in Figure \ref{figg1b}), predictive uncertainty estimation became possible, i.e. we were able to construct predictive envelopes that satisfy our desire in representing new target points probabilistically. In other words, using the BVM regression framework gives the user more control over the predictive envelopes and what uncertainties they represent.

\subsubsection{Numerical Example 3: Energy Dissipation Model\label{example3}}
\begin{figure} [H]
\centering
\includegraphics[width=10cm]{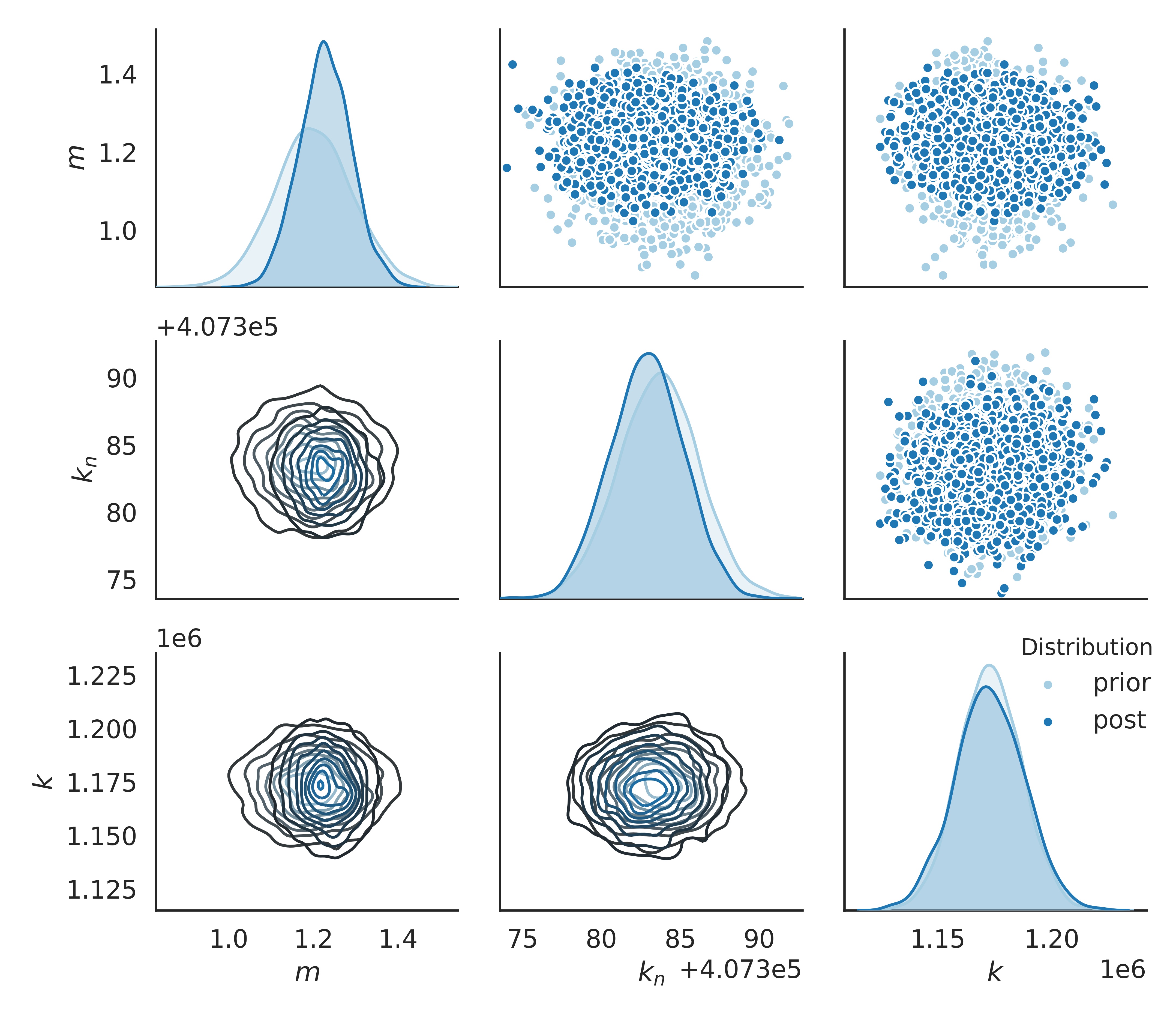}
\caption{Posterior inferences of model parameters.\label{corrplot}}
\end{figure}

\begin{figure} [H]
\centering
\includegraphics[width=10cm]{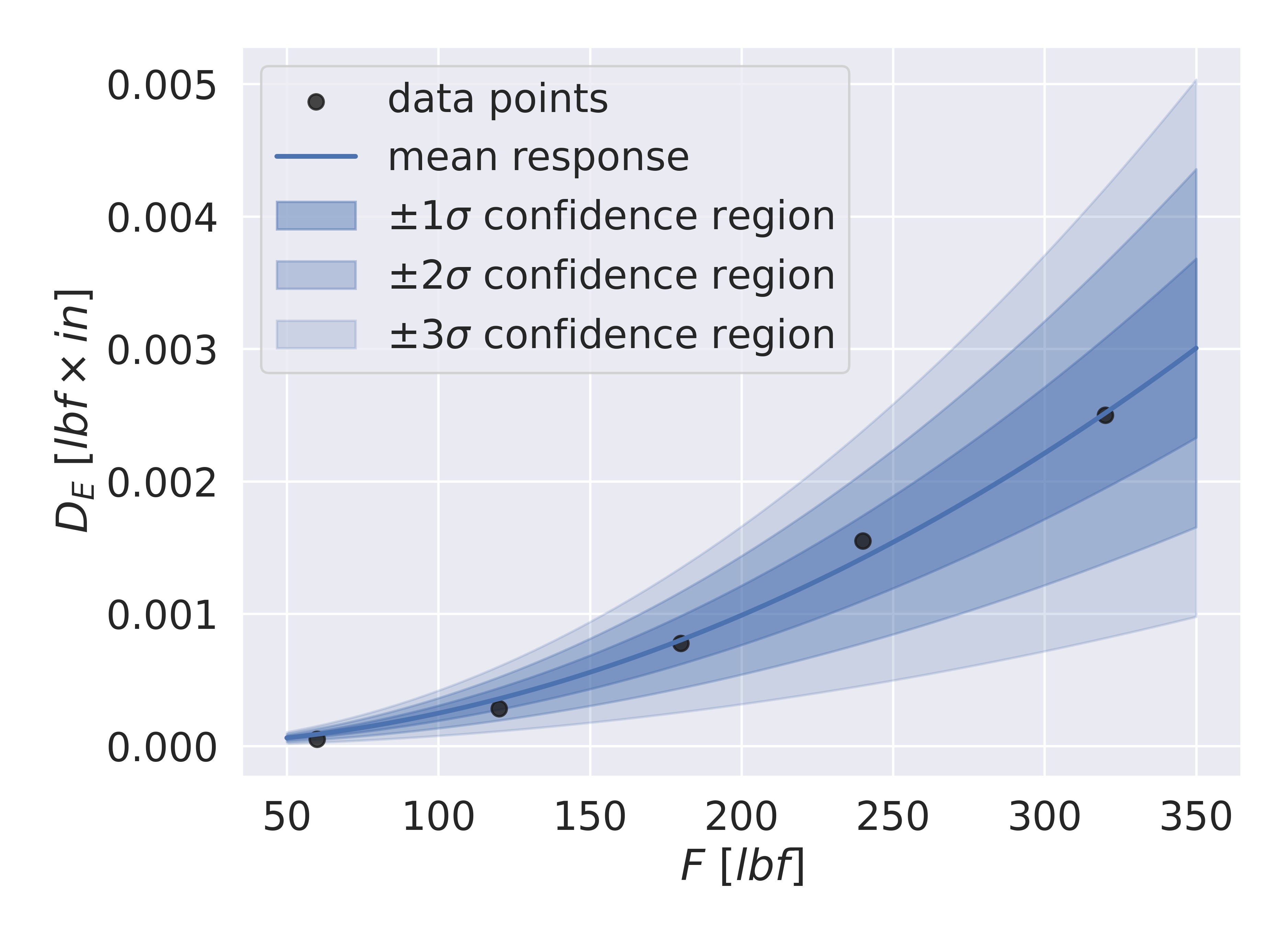}
\caption{Model response and predictive envelopes.\label{energyplot}}
\end{figure}

In this example, we consider the calibration of a three-parameter Smallwood model \cite{Rebba, Sankararaman2, sankararaman2012, smallwood2001, urbina2003} in order to predict the energy dissipation $D_E$ per cycle at a lap joint when subjected to an impact harmonic force of amplitude $F$. The energy loss in the joint under one cycle of sinusoidal loading is found by integrating the area under the hysteresis curve (force vs. displacement graph) and analytically derived as\vspace{-15pt}\\
\begin{align}
    D_E = k_n\bigg(\frac{m-1}{m+1}\bigg)\Delta z^{m+1}\label{energymodel}
\end{align}
where $k_n$ is a nonlinear stiffness, $m$ is a nonlinear exponent, and $\Delta z$ is the displacement amplitude obtained by solving\vspace{-15pt}\\
\begin{align}
    2F = k\Delta z - k_n\Delta z^m\label{displacementeq}
\end{align}
where $k$ is a linear stiffness term. The aim is to calibrate the three parameters $k_n$, $m$ and $k$ using the available input-output data, where the input corresponds to the force $F$ and the output corresponds to the measured dissipated energy $D_E$. Five levels of loading (and hence five measurements) were considered in the experiment and are shown in Table \ref{t3} below.
\begin{table}[H]
\begin{center}
 \begin{tabular}{ c c } 
 \toprule
Force  $F\:(lbf)$ & Energy $D_E\:(lbf \times in)$ \\ [0.2ex] 
 \midrule\vspace{-14pt}\\
60 & $5.30 \times 10^{-5}$ \\
120 & $2.85 \times 10^{-4}$ \\
180 & $7.78 \times 10^{-4}$ \\
240 & $1.55 \times 10^{-3}$ \\
320 & $2.50 \times 10^{-3}$ \\
 \bottomrule
\end{tabular}
\vspace*{5pt}
\caption{Smallwood model: Calibration data.\label{t3}}
\end{center}
\end{table}
We now follow the calibration procedure outlined in Section \ref{section3.2}. The calibration data is available in Table \ref{t3} and we have $\{X,Y\} = \{F, D_E\}$. The model $M$ can be expressed as $\hat{y} = D_E = M(F; k_n, m, k)$ which is obtained by combining Equations (\ref{energymodel}) and (\ref{displacementeq}) (by eliminating $\Delta z$). The prior distributions of the parameters $(m, k_n, k)$ are assumed to be Gaussian with the following mean and standard deviation.  
\begin{table}[H]
\begin{center}
 \begin{tabular}{ l c c c } 
 \toprule
Parameter & $m$ & $\log_{10}(k_n)$ & $k$\\ [0.2ex] 
 \midrule\vspace{-14pt}\\
Mean & $1.20$ & $5.61$ & 1172700 \\
Standard Deviation & $0.09$ & $0.40$ & 13760 \\
 \bottomrule
\end{tabular}
\vspace*{5pt}
\caption{Prior mean and standard deviation of model parameters.\label{t4}}
\end{center}
\end{table}
\noindent\vspace{-30pt}\\
Note that $m$ has no unit, $k$ and $k_n$ have units of $lbf/in$. We choose the comparison quantities to be the sets of model outputs and observed data, i.e. $\hat{z} \rightarrow \hat{Y}$ and $z \rightarrow Y$, and consider the $\epsilon-$Boolean agreement function (i.e. the improved reliability metric is included in our agreement definition) with a tolerance $\epsilon = 10^{-3}$. We assume the data to be completely certain (deterministic) and thus we use the likelihood function derived in (\ref{certainbvm}). We run MCMC for 10000 iterations with $20\%$ burn-in and we find that our method is able to construct the posterior inferences of the model parameters. The prior versus posterior distributions and the pairwise relationships between model parameters (using 3000 samples) are shown in Figure \ref{corrplot}.
From the posterior uncertainties of model parameters, we show the propagation of these uncertainties to the model response. In other words, we plot the predictive response and envelopes of the energy dissipation model, as shown in Figure \ref{energyplot}. 
Notice that the model (mean) response fits well the observed data, and the $68\%$ confidence band (i.e. $\pm 1\sigma$ confidence region) captures all of the data points. It is worth noting that in \cite{sankararaman2012}, the objective was to calibrate the nonlinear stiffness $k_n$, while treating the linear stiffness $k$ and the exponent $m$ as inputs (along with the force $F$). However, in our calibration process, we learned all three parameters ($m$, $k_n$, $k$) assuming $F$ to be the only input. From our previous discussions, we know that that the parameters' posterior inferences as well as the model's predictive envelopes are affected by the tolerance level $\epsilon$. Therefore, the analyst can always tune $\epsilon$ to produce predictive envelopes that better represent the observed data, resulting in a more reliable model. Since in this example we are using the $\epsilon$-Boolean agreement function, we know that the improved reliability metric \cite{Sankararaman2} is automatically taken into account in the calibration process.\\

To validate our model and test its reliability, we simply compute the BVM probability of agreement, $\mathcal{Z}(B)$ in Equation (\ref{BVMEvidence2}), using the available data and the posterior distributions of the model parameters. It is worth noting that in this case $\mathcal{Z}(B)$ is a measure of the model reliability. Using the posterior inferences of the model parameters, we calculate $\mathcal{Z}(B)$ to be 0.93. Therefore, given that this value is a measure of probability of agreement between the model outputs and the observed data, we can conclude that our model is acceptable according to the agreement function and the assumptions under consideration. Also note that our method is suitable for model selection; one can calibrate two models and then compute the BVM ratio (in Equation \ref{bvmratio} as described in Section \ref{section2.3}) to select the more reliable model.\\

Finally, remember that the reliability metric can always be a part of the agreement function (by combining it with the Boolean function that the analyst decides to work with, resulting in a compount Boolean as in the previous example). Also note that the BVM framework can be used in both the calibration stage as well as the validation stage, as we have shown above (which means that the reliability of the model can be taken into account in both stages).

\subsection{Discussion\label{discussion}}
As we have shown in the numerical examples above, the results are sensitive to the tolerance $\epsilon$. This arises from the fact that $\epsilon$ represents the modeler's tolerance on the difference between the model outputs and the observed data. Thus, any other parameter that is included in the Boolean function $B$ defined by the modeler will have an effect on the parameters' posterior distributions and the predictive envelopes. As shown in the previous examples, we can take advantage of this feature for modeling.\\

We analyse the effect of the tolerance $\epsilon$ as follows. Changing the agreement tolerance $\epsilon$ affects the acceptance rate in the MCMC iterations, i.e. a larger tolerance yields a larger variance of accepted ``candidate'' samples. The increased variance in the accepted samples produces wider posterior model parameter distributions. A smaller tolerance implies the converse.\\

Parameters not regressed inside the Boolean function $B$ ($\epsilon$ in this case) play a role similar to hyperparameters in Machine Learning.  By tuning the hyperparameters, the modeler can make the predictive envelopes more representative of the data, and improve the overall performance of the model when compared to a randomly selected test (or validation) data set. \\

Note that in the case of the $\epsilon$--Boolean, the user can increase $\epsilon$ indefinitely and still ``regress'' the model, although the predictive envelopes will be much wider than the data spread and hence less representative of the data. While widening the predictive envelopes can be useful for reliability and safety, if they are widened too much, the model can lose some of its predictive utility. To balance the trade-off between safety and utility, we regressed the model in (\ref{model2}) with respect to the compound Boolean (\ref{compbool}) in Section \ref{predex}. This Boolean function forced the model to be regressed in such a way that 95\%$\pm$4\% of the (uncertain) data points lie within the model's 95\% confidence region while simultaneously satisfying the $\langle\epsilon\rangle$ requirement (for diversity we used $\langle\epsilon\rangle$ instead of $\epsilon$ although nothing prevents us from doing so in principle). As stated in the caption of Figure \ref{BVMvsBMT}, we gradually tuned the value of $\langle\epsilon\rangle$ (hyperparameter tuning) to simultaneously satisfy both requirements imposed by the compound Boolean function while getting an accurate predictive uncertainty estimation.

\section{Conclusion\label{conclusion}}
In the calibration stage of model development, we can use the BVM to perform regression and model learning on data with any type of uncertainty, generate posterior parameter distributions, and model predictive envelopes, according to user-specified definitions of model-data agreement. The BVM regression framework proved its potential in offering new insights into the interpretation of the predictive envelopes of the Bayesian regression, standard regression, and likelihood-based techniques, and hence providing the analyst with more freedom and control over the predictive envelopes and their meaning. By calibrating models with respect to the validation metrics one desires a model to ultimately pass, reliability and safety metrics were integrated into our examples and automatically adopted by the model in the calibration phase. Finally, we find BVM regression to be a generalized regression and model learning tool allowing us to address several potential shortcomings in Bayesian and standard regression methods. \\

It is worth noting that all the work presented in this paper relies solely on parametric models. Thus, our method is limited to the cases and scenarios where an explicit expression of the model under question in terms of the parameters is known. Our future work involves expanding the BVM framework to nonparametric modeling (i.e. without making any assumptions about the model's expression). Finally, we emphasize the importance of practicing statistical responsibility by being explicit in the definition of agreement (between the model output and the data), the assumptions made, and the criteria considered when performing model calibration.

\paragraph{\textbf{Acknowledgments}}
This work was supported by the Center for Complex Systems (CCS) at King Abdulaziz City for Science and Technology (KACST) and the Massachusetts Institute of Technology (MIT). We would like to thank all the researchers in the CCS.
\bibliographystyle{unsrt}  
\bibliography{template}  

\appendix 

\section{Bayesian Model Testing\label{appendixA}}
We derive Equations (\ref{normalbmt}) -- (\ref{certainbmt}) mentioned in Section \ref{section2.2}. In the Bayesian model testing framework, the model output and the observed data are defined to agree only if their values are exactly equal. Thus, Bayesian model testing is a special case of the BVM where the agreement kernel is equal to the kronecker delta function (exact agreement) with continuous indices, i.e. $\Theta\big(B(\hat{Y},Y)\big) = \delta_{\hat{Y},Y} = \prod_{i=1}^n \delta_{\hat{y}_i,y_i}$. Since Bayesian model testing deals with probability densities, we have the following expression for the probability density of agreement (\ref{bvm}):\vspace{-10pt}\\
\begin{align}
\rho(A|M,D) = \displaystyle \frac{p(A|M,D)}{dA} &= \displaystyle\frac{1}{dA} \int_{\hat{Y},Y} \rho(\hat{Y}|M,D)\cdot\Theta\big(B(\hat{Y},Y)\big)\cdot\rho(Y|D)d\hat{Y}dY\notag\vspace{5pt}\\[\medskipamount]
&= \displaystyle\int_{\hat{Y},Y} \rho(\hat{Y}|M,D)\cdot\frac{\delta_{\hat{Y},Y}}{dA}\cdot\rho(Y|D)d\hat{Y}dY.\notag
\end{align}
The kronecker delta $\delta_{\hat{Y},Y}$ and the dirac delta $\delta(\hat{Y}-Y)$ functions are related as follows: \vspace{0pt}\\
\centerline{$\displaystyle \frac{\delta_{\hat{Y},Y}}{dA} = \delta(\hat{Y}-Y) = \prod_{i=1}^n\delta(\hat{y}_i-y_i).$}\vspace{5pt}\\
\noindent Thus, the probability density of agreement becomes,\vspace{-10pt}\\
\begin{align}
\rho(A|M,D) &= \displaystyle \int_{\hat{Y},Y} \rho(\hat{Y}|M,D)\cdot\delta(\hat{Y} - Y)\cdot\rho(Y|D)d\hat{Y}dY\notag\vspace{5pt}\\[\medskipamount]
&= \displaystyle \int_{\hat{Y},Y}\bigg(\int_{\vec{\alpha}}\rho(\hat{Y}|\vec{\alpha},M)\rho(\vec{\alpha}|M)d\vec{\alpha}\bigg)\cdot\delta(\hat{Y} - Y)\cdot\rho(Y|D)d\hat{Y}dY\notag\vspace{5pt}\\[\medskipamount]
&= \displaystyle\int_{\vec{\alpha}}\bigg( \int_{\hat{Y},Y}\rho(\hat{Y}|\vec{\alpha},M)\cdot\delta(\hat{Y} - Y)\cdot\rho(Y|D)d\hat{Y}dY\bigg)\cdot\rho(\vec{\alpha}|M)d\vec{\alpha}\notag\vspace{5pt}\\[\medskipamount]
&= \displaystyle\int_{\vec{\alpha}}\bigg( \int_{\hat{Y},Y}\underbrace{\delta\big(\hat{Y} - M(X;\vec{\alpha})\big)}_{\displaystyle \hat{Y} = M(X;\vec{\alpha})}\cdot\,\delta(\hat{Y} - Y)\cdot\rho(Y|D)d\hat{Y}dY\bigg)\cdot\rho(\vec{\alpha}|M)d\vec{\alpha}\notag\vspace{5pt}\\[\medskipamount]
&= \displaystyle\int_{\vec{\alpha}}\bigg( \int_{Y}\delta\big(M(X;\vec{\alpha}) - Y\big)\cdot\rho(Y|D)dY\bigg)\cdot\rho(\vec{\alpha}|M)d\vec{\alpha}.\label{appceq1}
\end{align}
\subsection{Normally Distributed Data}
If we assume the data to be normally distributed, i.e. $Y \sim \mathcal{N}(D,\Delta)$, we get, \notag\vspace{7pt}\\
\centerline{$\rho(Y|D) = \displaystyle \frac{1}{\sqrt{(2\pi)^n|\Delta|}}e^{\textstyle-\frac{1}{2}(Y - D)^T\Delta^{-1}(Y - D)},$}\vspace{7pt}\\
where $n$ is the dimension of the training data set, $\Delta$ is the covariance matrix, and $D$ is the observed data values. \vspace{5pt}\\
Therefore, using (\ref{appceq1}), we have,\vspace{-15pt}\\
\footnotesize
\begin{align}
\indent\hspace{-14pt}
\rho(A|M,D) &=  \displaystyle\int_{\vec{\alpha}}\bigg( \int_{Y}\delta\big(M(X;\vec{\alpha}) - Y\big)\cdot\displaystyle \frac{1}{\sqrt{(2\pi)^n|\Delta|}}e^{\textstyle-\frac{1}{2}(Y - D)^T\Delta^{-1}(Y - D)}dY\bigg)\cdot\rho(\vec{\alpha}|M)d\vec{\alpha},\notag\vspace{50pt}\\[\bigskipamount]
&= \displaystyle\int_{\vec{\alpha}}\underbrace{\displaystyle \frac{1}{\sqrt{(2\pi)^n|\Delta|}}e^{\textstyle-\frac{1}{2}\big(M(X;\vec{\alpha}) - D\big)^T\Delta^{-1}\big(M(X;\vec{\alpha}) - D\big)}}_{\displaystyle \mathcal{L}(\vec{\alpha})}\cdot\underbrace{\vphantom{\displaystyle \frac{1}{\sqrt{(2\pi)^n|\Delta|}}e^{\textstyle-\frac{1}{2}\big(M(X;\vec{\alpha}) - D\big)^T\Delta^{-1}\big(M(X;\vec{\alpha}) - D\big)}}\rho(\vec{\alpha}|M)d\vec{\alpha}}_{\displaystyle \mathcal{\pi(\vec{\alpha})}d\vec{\alpha}}.
\end{align}
\normalsize
Therefore, the likelihood function to be used in the MCMC algorithm is\vspace{-10pt}\\
\begin{align}
\mathcal{L}(\vec{\alpha}) &= \displaystyle \frac{1}{\sqrt{(2\pi)^n|\Delta|}}e^{\textstyle-\frac{1}{2}\big(M(X;\vec{\alpha}) - D\big)^T\Delta^{-1}\big(M(X;\vec{\alpha}) - D\big)},
\end{align}
which is Equation (\ref{normalbmt}) presented in Section \ref{section2.2}.
\subsection{Uniformly Distributed Data}
We first note that \vspace{-20pt}\\
\begin{equation}
dY \displaystyle\equiv d^nY \equiv \prod_{j=1}^n dy_j, \qquad j = 1,\hdots,n.\tag{C.2}\label{C.2.1}
\end{equation}
Now, we assume the data to be uniformly distributed, i.e.
\vspace{8pt}\\
\centerline{$y_j \sim \mathcal{U}(a_j,b_j), \qquad j = 1,\hdots,n.$}\vspace{8pt}\\
Then, the probability density $\rho(Y|D)$ becomes:\vspace{0pt}\\
\begin{equation}
\rho(Y|D) = \displaystyle \prod_{j=1}^n \rho(y_j|D) = \prod_{j = 1}^n \:\frac{\Theta\big(a_j\leq y_j \leq b_j\big)}{b_j - a_j}.\tag{C.3}\label{C.2.2}
\end{equation}
Notice that we can generalize $\rho(y_j|D) = \displaystyle\prod_{j = 1}^n\: \Theta\big(a_j\leq y_j \leq b_j\big)\rho(y_j|D)$ to any bounded probability density function (pdf). 
Therefore, using (\ref{appceq1}), we have,\vspace{-10pt}\\
\begin{align}
\indent\hspace{-20pt}
\rho(A|M,D) &=  \displaystyle\bigintssss_{\vec{\alpha}}\Bigg(\prod_{j=1}^n \bigintssss_{y_j}\delta\Big(M(x_j;\vec{\alpha}) - y_j\Big)\cdot\frac{\Theta\big(a_j\leq y_j \leq b_j\big)}{b_j - a_j}\:dy_j\Bigg)\cdot\rho(\vec{\alpha}|M)d\vec{\alpha},\notag\vspace{50pt}\\[\bigskipamount]
&= \displaystyle\bigintssss_{\vec{\alpha}}\underbrace{\displaystyle \Bigg(\prod_{j=1}^n\:\frac{\Theta\big(a_j\leq M(x_j;\vec{\alpha}) \leq b_j\big)}{b_j - a_j}\Bigg)}_{\displaystyle \mathcal{L}(\vec{\alpha})}\cdot\underbrace{\vphantom{\displaystyle \Bigg(\prod_{j=1}^n\:\frac{\Theta\big(a_j\leq M(x_j;\alpha) \leq b_j\big)}{b_j - a_j}\Bigg)}\rho(\vec{\alpha}|M)d\vec{\alpha}}_{\displaystyle \mathcal{\pi(\vec{\alpha})}d\vec{\alpha}}.
\end{align}
Therefore, the likelihood function to be used in the MCMC algorithm is\vspace{-10pt}\\
\begin{align}
\mathcal{L}(\vec{\alpha}) &= \displaystyle \prod_{j=1}^n\:\frac{\Theta\big(a_j\leq M(x_j;\vec{\alpha}) \leq b_j\big)}{b_j - a_j},
\end{align}
which is Equation (\ref{uniformbmt}) presented in Section \ref{section2.2}.
\subsection{Completely Certain Data}
If we consider the data to be completely certain, deterministic, i.e. $Y = D$, then, the probability density $\rho(Y|D)$ becomes $\rho(Y|D) = \delta(Y - D)$, and thus, using (\ref{appceq1}), we have,\vspace{-10pt}\\
\begin{align}
\rho(A|M,D) &=  \displaystyle\int_{\vec{\alpha}}\bigg( \int_{Y}\delta\big(M(X;\vec{\alpha}) - Y\big)\cdot\delta(Y - D)dY\bigg)\cdot\rho(\vec{\alpha}|M)d\vec{\alpha}\notag\vspace{50pt}\\[\bigskipamount]
&= \displaystyle\int_{\vec{\alpha}}\underbrace{\delta\big(M(X;\vec{\alpha}) - D\big)}_{\displaystyle \mathcal{L}(\vec{\alpha})}\cdot\underbrace{\vphantom{\delta\big(M(X;\vec{\alpha}) - D\big)}\rho(\vec{\alpha}|M)d\vec{\alpha}}_{\displaystyle \mathcal{\pi(\vec{\alpha})}d\vec{\alpha}}.
\end{align}
Therefore, the likelihood function to be used in the MCMC algorithm is\vspace{-10pt}\\
\begin{align}
\mathcal{L}(\vec{\alpha}) &= \displaystyle \delta\big(M(X;\vec{\alpha}) - D\big),
\end{align}
which is Equation (\ref{certainbmt}) presented in Section \ref{section2.2}.

\section{BVM Model Selection\label{appendixB}}
We derive Equations (\ref{BVMEvidence2}) -- (\ref{certainbvm}) presented in Section \ref{bvmregression}. We show how we can apply Bayesian model selection on any data distribution using the BVM probability of agreement. Starting from the original definition of probability of agreement (\ref{bvm}), we have,\vspace{-10pt}\\
\small
\begin{align}
p(A|M,D,B) &= \displaystyle \int_{\hat{Y},Y} \rho(\hat{Y}|M,D)\cdot\Theta\big(B(\hat{Y},Y)\big)\cdot\rho(Y|D)d\hat{Y}dY\notag\vspace{5pt}\\[\medskipamount]
&= \displaystyle \int_{\hat{Y},Y}\bigg(\int_{\vec{\alpha}}\rho(\hat{Y}|\vec{\alpha},M)\rho(\vec{\alpha}|M)d\vec{\alpha}\bigg)\cdot\Theta\big(B(\hat{Y},Y)\big)\cdot\rho(Y|D)d\hat{Y}dY\notag\vspace{5pt}\\[\medskipamount]
&= \displaystyle\int_{\vec{\alpha}}\bigg( \int_{\hat{Y},Y}\rho(\hat{Y}|\vec{\alpha},M)\cdot\Theta\big(B(\hat{Y},Y)\big)\cdot\rho(Y|D)d\hat{Y}dY\bigg)\cdot\rho(\vec{\alpha}|M)d\vec{\alpha}\notag\vspace{5pt}\\[\medskipamount]
&= \displaystyle\int_{\vec{\alpha}}\bigg( \int_{\hat{Y},Y}\delta\big(\hat{Y} - M(X;\vec{\alpha})\big)\cdot\Theta\big(B(\hat{Y},Y)\big)\cdot\rho(Y|D)d\hat{Y}dY\bigg)\cdot\rho(\vec{\alpha}|M)d\vec{\alpha}\notag\vspace{5pt}\\[\medskipamount]
&= \displaystyle\int_{\vec{\alpha}}\bigg( \int_{Y}\Theta\Big(B\big(M(X;\vec{\alpha}),Y\big)\Big)\cdot\rho(Y|D)dY\bigg)\cdot\rho(\vec{\alpha}|M)d\vec{\alpha},
\end{align}
\normalsize
which is Equation (\ref{BVMEvidence2}) derived in Section \ref{bvmregression}.
From (\ref{C.2.1}), we know that \vspace{5pt}\\
\centerline{$dY \displaystyle\equiv d^nY \equiv \prod_{j=1}^n dy_j, \qquad j = 1,\hdots,n.$}\vspace{5pt}\\
The probability density $\rho(Y|D)$ can be expressed as:\vspace{8pt}\\
\centerline{$\rho(Y|D) = \displaystyle \prod_{j=1}^n \rho(y_j|D).$}\vspace{5pt}\\
\noindent We also note that the compound Boolean under question can be expressed as:\vspace{8pt}\\
\centerline{$\Theta\Big(B\big(M(X;\vec{\alpha}),Y\big)\Big) = \displaystyle \prod_{j=1}^n \Theta\Big(B\big(M(x_j;\vec{\alpha}),y_j\big)\Big) \qquad j = 1,\hdots,n.$}\vspace{8pt}\\
Thus, we rewrite the BVM probability of agreement as follows:\vspace{10pt}\\
\centerline{$p(A|M,D,B) = \displaystyle\bigintssss_{\vec{\alpha}}\Bigg(\prod_{j=1}^n \bigintssss_{y_j}\Theta\Big(B\big(M(x_j;\vec{\alpha}),y_j\big)\Big)\cdot\rho(y_j|D)\:dy_j\Bigg)\cdot\rho(\vec{\alpha}|M)d\vec{\alpha}.$}\vspace{10pt}\\
We will use the $\epsilon-$Boolean indicator function defined as:\vspace{10pt}\\
\indent\centerline{$\Theta\Big(B\big(M(x_j;\vec{\alpha}),y_j\big)\Big)$= \begin{math}
    \begin{dcases}
        1, & \text{if } \big|y_j - M(x_j;\vec{\alpha})\big| \displaystyle\leq \epsilon \\[\smallskipamount]
        0, & \text{otherwise}\\
    \end{dcases}
\end{math}}\vspace{8pt}\\
where $j = 1,\hdots,n$.\\
Then, the indicator function can be rewritten as:\vspace{-10pt}\\
\begin{align}
\Theta\Big(B\big(M(x_j;\vec{\alpha}),y_j\big)\Big) &= \Theta\Big(\big|y_j - M(x_j;\vec{\alpha})\big| \displaystyle\leq \epsilon\Big)\notag\\[\medskipamount]
&= \Theta\Big(M(x_j;\vec{\alpha}) - \epsilon \leq y_j \leq M(x_j;\vec{\alpha}) + \epsilon\Big),\notag
\end{align}
where $j = 1,\hdots,n$. Therefore, the BVM probability of agreement can be expressed as:\vspace{-5pt}\\
\footnotesize
\begin{equation}
p(A|M,D,B) = \displaystyle\bigintssss_{\vec{\alpha}}\Bigg(\prod_{j=1}^n \bigintssss_{y_j}\Theta\Big(M(x_j;\vec{\alpha}) - \epsilon \leq y_j \leq M(x_j;\vec{\alpha}) + \epsilon\Big)\cdot\rho(y_j|D)\:dy_j\Bigg)\cdot\rho(\vec{\alpha}|M)d\vec{\alpha}.\tag{D.1}\label{D.1}
\end{equation}
\normalsize
\subsection{Normally Distributed Data\label{appendixB1}}
Note that the Boolean $\Theta\Big(M(x_j;\vec{\alpha}) - \epsilon \leq y_j \leq M(x_j;\vec{\alpha}) + \epsilon\Big)$ is equal to 1 only when $y_j$ belongs to the interval $\Big[M(x_j;\vec{\alpha}) - \epsilon, \:M(x_j;\vec{\alpha}) + \epsilon\Big]$.\vspace{10pt}\\
Thus, the BVM probability of agreement becomes:\vspace{-15pt}\\
\begin{align}
p(A|M,D,B) &= \displaystyle\bigintssss_{\vec{\alpha}}\Bigg(\prod_{j=1}^n \bigintssss_{\:M(x_j;\vec{\alpha}) - \epsilon}^{\:M(x_j;\vec{\alpha}) + \epsilon}\rho(y_j|D)\:dy_j\Bigg)\cdot\rho(\vec{\alpha}|M)d\vec{\alpha}.\notag
\end{align}
\noindent Now, we assume that the data is normally distributed, i.e.
\vspace{8pt}\\
\centerline{$y_j \sim \mathcal{N}(D_j,\sigma^2_j), \qquad j = 1,\hdots,n.$}\vspace{10pt}\\
Then, the probability density $\rho(Y|D)$ becomes:\vspace{2pt}\\
\centerline{$\rho(Y|D) = \displaystyle \prod_{j=1}^n \rho(y_j|D) = \prod_{j = 1}^n \frac{1}{\sqrt{2\pi}\sigma_j}\text{e}^{-\textstyle\frac{1}{2}\Big(\frac{y_j - D_j}{\sigma_j}\Big)^2}.$}\vspace{8pt}\\
Thus, we rewrite the BVM probability of agreement as follows:\vspace{-10pt}\\
\small
\begin{align}
p(A|M,D,B) &= \displaystyle\bigintssss_{\vec{\alpha}}\Bigg(\prod_{j=1}^n \bigintssss_{\:M(x_j;\vec{\alpha}) - \epsilon}^{\:M(x_j;\vec{\alpha}) + \epsilon}\rho(y_j|D)\:dy_j\Bigg)\cdot\rho(\vec{\alpha}|M)d\vec{\alpha}\notag\\[\bigskipamount]
&= \displaystyle\bigintssss_{\vec{\alpha}}\Bigg(\prod_{j=1}^n \bigintssss_{\:M(x_j;\vec{\alpha}) - \epsilon}^{\:M(x_j;\vec{\alpha}) + \epsilon}\frac{1}{\sqrt{2\pi}\sigma_j}\text{e}^{-\textstyle\frac{1}{2}\Big(\frac{y_j - D_j}{\sigma_j}\Big)^2}\:dy_j\Bigg)\cdot\rho(\vec{\alpha}|M)d\vec{\alpha}\notag\\[\bigskipamount]
&= \displaystyle\bigintssss_{\vec{\alpha}}\underbrace{\:\displaystyle\prod_{j=1}^n\Bigg(F\Big(M(x_j;\vec{\alpha}) + \epsilon\Big) - F\Big(M(x_j;\vec{\alpha}) - \epsilon\Big)\Bigg)}_{\displaystyle \mathcal{L}(\vec{\alpha},B)}\cdot\underbrace{\vphantom{\prod_{j=1}^n\Bigg(\frac{u_j - l_j}{b_j - a_j}\Bigg)}\rho(\vec{\alpha}|M)d\vec{\alpha}}_{\displaystyle \mathcal{\pi(\vec{\alpha})}d\vec{\alpha}},
\end{align}
\normalsize
where $F(x)$ is the cumulative distribution function (cdf) expressed as:\vspace{8pt}\\
\centerline{$F(x) = \displaystyle\Phi\bigg(\frac{x - D}{\sigma}\bigg)$,}\vspace{10pt} 
where $\Phi(\cdot)$ is the cumulative distribution function of the standard normal distribution, i.e. $\mathcal{N}(0,1)$, and expressed as:\vspace{5pt}\\
\centerline{$\Phi(x) = \displaystyle \frac{1}{\sqrt{2\pi}}\bigintssss_{-\infty}^{\textstyle x}\text{e}^{\textstyle-\frac{t^2}{2}}\:dt$.}\vspace{20pt}\\
Therefore, the likelihood function to be used in the MCMC algorithm is\vspace{-10pt}\\
\begin{align}
\mathcal{L}(\vec{\alpha},B) &= \displaystyle\prod_{j=1}^n\Bigg(F\Big(M(x_j;\vec{\alpha}) + \epsilon\Big) - F\Big(M(x_j;\vec{\alpha}) - \epsilon\Big)\Bigg).
\end{align}
\subsection{Uniformly Distributed Data}
If we assume the data to be uniformly distributed (\ref{C.2.2}), then the BVM probability of agreement (\ref{D.1}) becomes:\vspace{-10pt}\\
\footnotesize
\begin{align}
p(A|M,D,B) = \displaystyle\bigintssss_{\vec{\alpha}}\Bigg(\prod_{j=1}^n \bigintssss_{y_j}\Theta\Big(M(x_j;\vec{\alpha}) - \epsilon \leq y_j \leq M(x_j;\vec{\alpha}) + \epsilon\Big)\cdot\frac{\Theta\big(a_j\leq y_j \leq b_j\big)}{b_j - a_j}\:dy_j\Bigg)\cdot\rho(\vec{\alpha}|M)d\vec{\alpha}.
\end{align}
\normalsize
Note that the product $\Theta\Big(M(x_j;\vec{\alpha}) - \epsilon \leq y_j \leq M(x_j;\vec{\alpha}) + \epsilon\Big)\cdot \Theta\Big(a_j\leq y_j \leq b_j\Big)$ is equal to 1 only when $y_j$ belongs to both intervals $\Big[M(x_j;\vec{\alpha}) - \epsilon, \:M(x_j;\vec{\alpha}) + \epsilon\Big]$ and $\Big[a_j, \:b_j\Big]$.\vspace{10pt}\\
\noindent Let $l_j$ and $u_j$ be such that\vspace{11pt}\\
\centerline{$\Big[l_j,\: u_j\Big] = \Big[M(x_j;\vec{\alpha}) - \epsilon,\: M(x_j;\vec{\alpha}) + \epsilon\Big] \cap \Big[a_j,\: b_j\Big] \qquad j=1,\hdots,n$}\vspace{12pt}\\
Thus, the BVM probability of agreement becomes:\vspace{-10pt}\\
\begin{align}
p(A|M,D,B) &= \displaystyle\bigintssss_{\vec{\alpha}}\Bigg(\prod_{j=1}^n \bigintssss_{y_j}\frac{\Theta\big(l_j\leq y_j \leq u_j\big)}{b_j - a_j}\:dy_j\Bigg)\cdot\rho(\vec{\alpha}|M)d\vec{\alpha}\notag\\[\bigskipamount]
&= \displaystyle\bigintssss_{\vec{\alpha}}\Bigg(\prod_{j=1}^n \bigintssss_{\:l_j}^{\:u_j}\frac{1}{b_j - a_j}\:dy_j\Bigg)\cdot\rho(\vec{\alpha}|M)d\vec{\alpha}\notag\\[\bigskipamount]
&= \displaystyle\bigintssss_{\vec{\alpha}}\underbrace{\Bigg(\prod_{j=1}^n\:\frac{u_j - l_j}{b_j - a_j}\Bigg)}_{\displaystyle \mathcal{L}(\vec{\alpha},B)}\cdot\underbrace{\vphantom{\Bigg(\prod_{j=1}^n\:\frac{u_j - l_j}{b_j - a_j}\Bigg)}\rho(\vec{\alpha}|M)d\vec{\alpha}}_{\displaystyle \mathcal{\pi(\vec{\alpha})}d\vec{\alpha}}.
\end{align}
Therefore, the likelihood function to be used in the MCMC algorithm is\vspace{-10pt}\\
\begin{align}
\mathcal{L}(\vec{\alpha},B) &= \prod_{j=1}^n\:\frac{u_j - l_j}{b_j - a_j},\notag 
\end{align}
which is Equation (\ref{uniformbvm}) presented in Section \ref{bvmregression}.
\subsection{Completely Certain Data}
If we consider the data to be completely certain, deterministic, i.e. $Y = D$, then, the probability density $\rho(Y|D)$ becomes $\rho(Y|D) = \delta(Y - D)$, and thus, using (\ref{BVMEvidence2}), we have,\vspace{-10pt}\\
\begin{align}
p(A|M,D,B) &=  \displaystyle\int_{\vec{\alpha}}\bigg( \int_{Y}\Theta\Big(B\big(M(X;\vec{\alpha}),Y\big)\Big)\cdot\delta(Y - D)dY\bigg)\cdot\rho(\vec{\alpha}|M)d\vec{\alpha}\notag\vspace{50pt}\\[\bigskipamount]
&= \displaystyle\int_{\vec{\alpha}}\underbrace{\Theta\Big(B\big(M(X;\vec{\alpha}),D\big)\Big)}_{\displaystyle \mathcal{L}(\vec{\alpha},B)}\cdot\underbrace{\vphantom{\Theta\Big(B\big(M(x;\vec{\alpha}),D\big)\Big)}\rho(\vec{\alpha}|M)d\vec{\alpha}}_{\displaystyle \mathcal{\pi(\vec{\alpha})}d\vec{\alpha}}.
\end{align}
Therefore, the likelihood function to be used in the MCMC algorithm is\vspace{-10pt}\\
\begin{align}
\mathcal{L}(\vec{\alpha},B) &= \Theta\Big(B\big(M(X;\vec{\alpha}),D\big)\Big),
\end{align}
which is Equation (\ref{certainbvm}) presented in Section \ref{bvmregression}.






\end{document}